\documentclass[aps,prb,showpacs,superscriptaddress]{revtex4}

\usepackage[dvips]{graphicx}

\begin{document}
%\draft
%\preprint{}

\title{Scaling analysis of electron transport through metal-semiconducting 
carbon nanotube interfaces:  Evolution from the molecular limit to the 
bulk limit}


%\affiliation can be followed by \email, \homepage, \thanks as well.
%\author{Yongqiang Xue}
%\thanks{}
\author{Yongqiang Xue$^{*}$}
\affiliation{College of Nanoscale Science and Engineering, University 
at Albany, State University of New York, Albany, New York 12203, USA}
\email{yxue@uamail.albany.edu}
\homepage{http://www.albany.edu/~yx152122}
\author{Mark A. Ratner}
\affiliation{Department of Chemistry and Materials Research Center, 
Northwestern University, Evanston, Illinois 60208, USA}

\date{\today }
%\maketitle

\begin{abstract}
We present a scaling analysis of electronic and transport properties of 
metal-semiconducting carbon nanotube interfaces as a function of the 
nanotube length within the coherent transport regime, which takes fully 
into account atomic-scale electronic structure and three-dimensional 
electrostatics of the metal-nanotube interface using a real-space Green's 
function based self-consistent tight-binding theory. As the first example, 
we examine devices formed by attaching finite-size single-wall carbon 
nanotubes (SWNT) to both high- and low- work function metallic electrodes 
through the dangling bonds at the end, where the length of the SWNT 
molecule varies from the molecular limit to the bulk limit and the strength of 
metal-SWNT coupling varies from the strong coupling to the weak coupling 
limit. We analyze the nature of Schottky barrier formation at the 
metal-nanotube interface by examining the electrostatics, the band lineup 
and the conductance of the metal-SWNT molecule-metal junction as a 
function of the SWNT molecule length and metal-SWNT coupling strength. 
We show that the confined cylindrical geometry and the atomistic nature 
of electronic processes across the metal-SWNT interface leads to a different 
physical picture of band alignment from that of the planar metal-semiconductor 
interface. We analyze the temperature and length dependence of the 
conductance of the SWNT junctions, which shows a transition from tunneling- 
to thermal activation-dominated transport with increasing nanotube length. 
The temperature dependence of the conductance is much 
weaker than that of the planar metal-semiconductor interface due to the finite 
number of conduction channels within the SWNT junctions. We find that 
the current-voltage characteristics of the metal-SWNT molecule-metal 
junctions are sensitive to models of the potential response to the applied 
source/drain bias voltages. Our analysis applies in general to devices based on 
quasi-one-dimensional nanostructures including molecules, carbon nanotubes 
and semiconductor nanowires. 
\end{abstract}

%\keywords{}

\maketitle

%\section{}
%\subsection{}
%\subsubsection{}

\section{Introduction}
It is interesting to note that all the semiconductor devices that have had a 
sustaining impact on integrated microelectronics were invented before 
1974,~\cite{Sze1} the year when Chang, Esaki and Tsu reported the first 
observation of negative differential resistance (NDR) in semiconductor 
heterojunction resonant-tunneling diodes (RTD).~\cite{RTD} The operation 
of such semiconductor devices relies on the (controlled) presence of 
imperfections in otherwise perfect crystals,~\cite{Shockley} through 
doping or through interfaces between materials with different electronic 
and/or lattice structures. Doping introduces electronic impurities 
(electrons/holes) into the otherwise perfect band structure through 
introducing atomic impuries (dopants) into the otherwise perfect lattice 
structure.~\cite{E-H} The presence of interfaces, on the other hand, 
induces spatial charge and potential inhomogeneities which control the 
injection and modulate the motion of excess charge carriers within 
the device. A number of fundamental buiding blocks of 
microelectronics can therefore be identified according to the interface 
structures that control the device operation,~\cite{Sze1,Sze2} including 
metal-semiconductor (MS) interfaces, semiconductor homo-(p-n) 
junctions, semiconductor heterojunctions and metal-insulator-semiconductor 
(MIS) interfaces.~\cite{Sze1,Sze2} Despite the continuous shrinking of feature 
size and correspondingly the increasing importance of hot-carrier and 
quantum mechanical effects,~\cite{HotQu} the design and operation of 
semiconductor transistors have followed remarkably well the scaling 
rules for device miniaturization~\cite{Scaling} derived from the semi-classical 
semiconductor transport equations.~\cite{DD,SeBook} There are also  
theoretical arguments that support the use of semiclassical pictures even in 
high-field transport~\cite{WilkSemi} and nanoscale 
ballistic silicon transistors.~\cite{Lundstrom}

The discovery of single-wall carbon nanotubes (SWNTs) in the early 
1990s~\cite{Iijima} has led to intense world-wide activity exploring their 
electrical properties and potential applications in nanoelectronic 
devices.~\cite{Dekker99,Dress98,NTDevice} SWNTs are nanometer-diameter 
all-carbon cylinders with unique structure-property relations: They consist of 
a single graphene sheet wrapped up to form a tube and their physical 
properties are controlled by the boundary conditions imposed on the 
wrapping directions. They provide ideal artifical laboratories for studying 
transport on the length scale ranging from the molecular limit as 
all-carbon cylindrical molecules to the bulk limit as quasi-one-dimensional 
quantum wires with the same lattice configuration and local bonding 
environment.~\cite{Dekker99,Dress98,NTDevice} Many device concepts 
well known in conventional semiconductor microelectronics have been 
successfully demonstrated on a single-tube basis, ranging from intramolecular 
homo(hetero)-junctions, modulation doping to  
field-effect transitors.~\cite{DekkerNT,AvNT,McEuNT,DaiNT,JohnNT} 
This prompts interest in knowing if the physical mechanisms underlying the 
operation of conventional microelectronic devices remain valid down to such 
ultra-small scales. Research on SWNT-based nanoelectronic devices therefore 
presents unique opportunities both for exploring novel device technology 
functioning at the nano/molecular-scale and for re-examining the physical 
principles of semiconductor microelectronics from the bottom-up atomistic 
approach. In addition, the concepts and techniques developed can be readily 
generalized to investigate other quasi-one-dimensional nanostructures, 
in particular semiconductor nanowires.~\cite{Wire}  

Among the device physics problems arising in this context, the nature of 
electron transport through a metal-semiconducting SWNT 
interface~\cite{Xue99NT,TersoffNT,Odin00,De02} stands out due 
to its simplicity and its role as one of the basic device building 
blocks.~\cite{Sze1,Sze2,MSBook}  As the device building block, it is also 
crucial for understanding the mechanisms and guiding the design of 
SWNT-based electrochemical sensors,~\cite{NTCH} electromechanical 
devices,~\cite{NTME} and field-effect transistors 
(NTFET),~\cite{AvFET,DaiFET,McEuFET} where electron transport through 
the metal-SWNT-metal junction is modulated through molecular adsorption, 
mechanical strain and electrostatic gate field respectively. 
Note that in the case of NTFET, metals have been used as the source, drain and 
gate electrodes, in contrast to silicon-based transistors which use 
heavily-doped polycrystalline materials.~\cite{AvFET,DaiFET,McEuFET} 

The nature of charge transport 
through metal-semiconductor interfaces has been actively investigated for 
decades due to their importance in microelectronic 
technology,~\cite{MSBook,MSMonch,MSReview} but is still not fully resolved, in 
particular regarding the mechanism of Schottky barrier formation/height 
and high-field transport phenomena.~\cite{MSPaper,MSTran} 
Compared to their bulk semiconductor counterpart, metal-SWNT interfaces 
present new challenges in that: (1) Both the contact area and the active 
device region can have atomic-scale dimensions; (2) The quasi-one-dimensional 
structure (cylindrical for nanotube materials) makes the screening of 
electron-electron interaction ineffective and leads to long range correlation 
between electrons within SWNT-based devices; (3) Last but probably the 
most important difference lies in the fact that quasi-one-dimensional wires, 
no matter how long, cannot be treated as electron reservoirs.~\cite{Landauer} 
This is partly due to the fact that the restricted phase space in such systems 
prevents rapid relaxation of injected carriers to a pre-defined equilibrium 
state through electron-electron and/or electron-phonon scatterings. But 
more importantly, this can be understood from a simple geometrical 
argument: Since the total current is conserved, there will always be a 
finite current density flowing along the wire and consequently a 
non-equilibrium state persists no matter how strong electron-electron 
and/or electron-phonon scattering is. An equilibrium state can be achieved 
only through the widened (adiabatic) contact with the (three-dimensional) 
metallic electrodes (or other macroscopic measurement apparatus) attached 
to them, where the finite current density can be effectively ``diluted'' 
through the larger cross sectional area.~\cite{Landauer} 
\emph{Correspondingly electron transport through metal-SWNT interfaces 
can only be studied within the configuration of metal-SWNT-metal junction} 
(as are other quasi-one-dimensional systems), in contrast to the planar 
metal-semiconductor interface, where the presence of the second electrode 
can be implicitly neglected and the analysis of transport characteristics 
proceeds by analyzing the interface region and the bulk semiconductor 
region separately.~\cite{MSBook} 

The last fact has important implications in the assessment of Schottky 
barrier effects on the measured transport characteristics, since 
transport mechanisms both at the interface and inside the active device 
region have to be considered simultaneously even for a long nanotube. 
Since the back-scattering of electrons by impurities~\cite{Ando} and 
the low-energy acoustic phonons~\cite{Kane,WT98,NTPhonon} are weak in 
such quasi-one-dimensional systems, the nature of the electron 
transport through metal-SWNT interfaces generally depends on the type 
of the SWNTs (length/diameter/chirality), the type of the contacts, and 
the temperature and bias voltage range. Experimentally, this matter is 
further complicated by the different fabrication/contact schemes used and 
the lack of knowledge of the atomic structure of the SWNT junctions. 
 
Recent works have studied electrical transport through a metal-long carbon 
nanotube interface using the bulk (infinitely long) band structures and 
electrostatics of ideal cylinders~\cite{TersoffNT,Odin00,KM01,De02}. 
For nanoelectronics research, it will be important to explore the device 
functionality of finite-size carbon nanotubes with lengths ranging from 
nanometers to tens of nanometers. Since most of the SWNT devices currently 
investigated are based on SWNTs with length of $100nm$ or longer, an 
investigation of the finite-size effect will shed light on the scaling limit of 
carbon nanotube devices,~\cite{Limit,NTLimit,GuoNT} as well as establish 
the validity or viability of using bulk device physics concepts in nanotube 
device research. 
 
The finite-size SWNT can be either a finite cylindrical all-carbon molecule 
attached to the metal surfaces through the dangling bonds at the end 
(end-contact scheme),~\cite{XueNT03} a finite segment of a long carbon 
nanotube wire whose ends are buried inside the metallic electrodes 
(embedded-contact scheme),~\cite{DaiNT,XueNT04} or a finite segment of 
a long nanotube wire which is deposited on top of predefined metallic 
electrodes and side-contacted to the surfaces of the electrodes 
(side-contact sheme).~\cite{Dekker99,Contact} In the case of finite SWNT 
molecules, a transition from the molecular limit to the bulk (infinitely long) 
limit in the electronic structure will occur as the length of the finite SWNT 
varies from nanometers to tens of nanometers. In the case of long SWNT 
wires, the electronic structure of the finite SWNT segment remains that 
of the bulk (which may be perturbed by the coupling to the electrodes), 
but the electrostatics of the metal-SWNT-metal junction varies with the 
SWNT length. Due to the nanoscale contact geometry and reduced 
dimensionality of SWNTs, a correct description of the Schottky barrier 
formation at the metal-finite SWNT interface generally requires an atomistic 
description of the electronic processes throughout the metal-SWNT-metal 
junctions. 

\begin{figure}
\includegraphics[height=3.2in,width=4.0in]{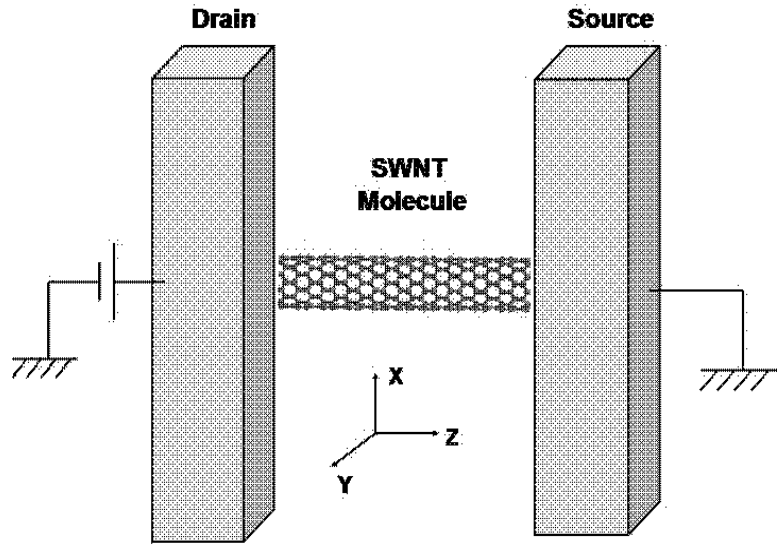}
\caption{\label{xueFig1} (Color online) Schematic illustration of the 
metal-SWNT molecule-metal junction. We have also shown the 
coordinate system of the SWNT junction. }
\end{figure}

The purpose of this work is thus to present a self-consistent atomistic 
analysis of the electronic and transport properties of the metal-SWNT 
interfaces within the configuration of metal-SWNT-metal junction as a 
function of the SWNT length, which is varied from the nanometer to tens 
of nanometer range. In contrast to previous theoretical 
works,~\cite{Xue99NT,TersoffNT,Odin00,KM01,De02} we use a novel Green's 
function based self-consistent tight-binding (SCTB) theory in real-space, 
which takes fully into account the three-dimensional electrostatics and 
the atomic-scale electronic structure of the SWNT junctions. In accordance 
with the nanometer length-scale of the SWNT studied, we treat electron 
transport within the coherent transport regime.~\cite{WT98,NTPhonon} 
In this first paper, we 
consider the device formed by attaching a finite cylindrical SWNT molecule 
to the electrode surface through the dangling bonds at the end 
(Fig.\ \ref{xueFig1}). The case of a finite-segment of long SWNT wires in 
both embedded-contact and side-contact schemes will be treated in the 
subsequent paper. 

The device configuration considered here 
represents an atomic-scale analogue (both the contact area and the active 
device region are atomic-scale) to the planar metal-semiconductor 
interface, where dangling bonds also exist at the semiconductor surface 
layers and contribute to the Schottky 
barrier formation.~\cite{MSMonch,MSReview}  
Compared to other molecular-scale devices where 
the individual organic molecule is self-assembled onto the metallic electrode 
through appropriate end groups~\cite{MEReview,XueMol1,XueMol2}, 
the SWNT molecule presents a homogeneous device structure where the only 
electronic inhomogeneity is introduced at the metal-SWNT interface through 
the ring of dangling-bond carbon atoms. The device structure considered 
here thus provides an ideal system for studying the length dependence 
of device characteristics on an atomic scale. In particular, the effect of the 
coupling strength can be studied by varing the SWNT end-electrode 
surface distance.  

The rest of the paper is organized as follows: We present the details of the 
Green's function based self-consistent tight-binding model in section II. We 
analyze the evolution of the SWNT electronic structure with the length of 
the SWNT molecules in section III. We devote section IV to analyzing the 
nature of Schottky barrier formation at the metal-SWNT molecule interface 
by examining the electrostatics (change transfer and electrostatic potential 
change), the electron transmission characteristics and the ``band'' lineup. 
In section V, we present the temperature and length dependence of the 
SWNT junction conductance. We show in section VI that the 
current-voltage (I-V) characteristics of the SWNT junction are sensitive 
to the spatial variation of the voltage drop across the junction. Finally 
in section VII, we summarize our results and discuss their implications 
for the functioning of SWNT-based devices. A preliminary report of some 
of the results presented here has been published 
elsewhere.~\cite{XueNT03} We use atomic units throughout the paper 
unless otherwise noted.

\section{Theoretical Model}

\subsection{\label{Theory} Real-space Green's function based 
self-consistent tight-binding (SCTB) theory}

Modeling electron transport in nanoscale devices is much more difficult than 
in bulk and mesoscopic semiconductor devices due to the necessity of 
including microscopic treatment of the electronic structure and the 
contacts to the measurement electrodes, which requires combining the 
non-equilibrium statistical mechanics of a open quantum 
system~\cite{MesoPhy,Buttiker86,LB88} with an 
atomistic modeling of the electronic structure.~\cite{Xue02,GuoAB}. For 
small molecular-scale devices where the inelastic carrier scattering can be 
neglected, this has been done using a self-consistent Matrix Green's function 
(SCMGF) method,~\cite{XueMol1,XueMol2,Xue02} which combines the 
Non-Equilibrium Green's Function (NEGF) theory of quantum 
transport~\cite{MesoEE,NEGF} with an effective single-particle description 
of the electronic structure using density-functional theory (DFT).~\cite{DFT} 
To treat larger nanoscale systems, e.g., carbon nanotubes or semiconductor 
nanowires containing thousands or tens of thousands of atoms, a simpler 
tight-binding-type theory is more 
appropriate.~\cite{TB,Lake97,Xue99Mol,JoachimTB,DattaMol,RatnerMol,Frauenheim} 
Correspondingly, we have developed a real-space self-consistent 
tight-binding (SCTB) method which includes atomic-scale description 
of the electronic structure and the three-dimensional electrostatics 
of the metal-SWNT-metal junction. The method is essentially the 
semi-empirical version of the SCMGF method for treating molecular 
electronic devices and is applicable to arbitrary nanostructured 
devices. The details and applications of the SCMGF method have been described 
extensively elsewhere~\cite{XueMol1,XueMol2,Xue02}, here we give a 
brief summary of the self-consistent tight-binding implementation. 

The method starts from the Hamiltonian $H_{0}$ describing the isolated 
nanostructure and the bare metallic electrodes, which can be obtained 
using either \emph{ab initio} or empirical approaches as appropriate. 
The effect of the coupling to the electrodes is included as self-energy 
operators.~\cite{MesoEE,NEGF} The coupling to the external contacts 
leads to charge transfer between the electrodes and the nanostructure. 
Applying a finite bias voltage also leads to charge redistribution (screening) 
within the nanostructure. Both the effect of the coupling to the contact and 
the screening of the applied field thus introduced will need to be treated 
self-consistently. The Hamiltonian describing the coupled 
metal-nanostructure-metal junction is thus 
$H=H_{0}+V_{ext}+\delta V[\delta \rho]$, where an external potential 
of the type $V_{ext}(\vec r)=-e\vec E \cdot \vec r$ should be added in the 
case of a nonzero source-drain or gate voltage and $\delta \rho$ is the 
change in the charge density distribution. Given the Hamiltonian matrix, the 
density matrix $\rho_{ij}$ and therefore the electron density are calculated 
using the Non-Equilibrium Green's Function (NEGF) 
method~\cite{Xue02,GuoAB,MesoEE,NEGF} 
from either 
\begin{eqnarray}
\label{GE}
G^{r} 
&=& \{ E^{+}S-H-\Sigma_{L}(E)-\Sigma_{R}(E) \}^{-1}, \\
\rho &=& \int \frac{dE}{2\pi }Imag[G^{r}](E).
\end{eqnarray}
for device at equilibrium or 
\begin{eqnarray}
\label{GNE}
G^{<} 
&=& i[G^{r}(E)\Gamma_{L}(E)G^{a}(E)]f(E-\mu_{L}) \\ \nonumber
&+& i[G^{r}(E)\Gamma_{R}(E)G^{a}(E)]f(E-\mu_{R}), \\
\rho &=& \int \frac{dE}{2\pi i}G^{<}(E).
\end{eqnarray}
for device at non-equilibrium.  Here $S$ is overlap matrix and 
$f(E-\mu_{L(R)})$ is the Fermi-Dirac distribution function at the left 
(right) electrode. The Green's functions $G^{r}$ and $G^{<}$ are 
defined in the standard manner.~\cite{NEGF,MesoEE} $\Sigma_{L(R)}$ 
is the self-energy operator due to the coupling to the left (right) electrode 
which is calculated from the metal surface Green's function, while 
$\Gamma_{L(R)}=i(\Sigma_{L(R)}-(\Sigma^{\dagger}_{L(R)})$ (See 
Refs.\ \onlinecite{Xue02,MesoEE} for details). 

Within the 
local-density-approximation of density-functional theory,~\cite{DFT} 
the long range part of $\delta V$ is just the coulomb potential 
$\delta V(\vec r)=\int \frac{\delta \rho (\vec r')}{|\vec r- \vec r'|}d\vec r'$. 
For self-consistent treatment of the charging effect within the tight-binding 
formulation, we follow the density-functional tight-binding (DFTB) theory 
developed by Frauenheim and coworkers~\cite{Frauenheim} by 
approximating the charge distribution as a superposition of normalized 
atomic-centered charge distributions 
$\delta \rho(\vec r)=\sum_{i} \delta N_{i} \rho_{i}(\vec r-\vec r_{i})$ 
where $\delta N_{i}$ is the net number of electrons on atomic-site $i$ and 
$\rho_{i}$ is taken as a normalized Slater-type function 
$\rho_{i}(\vec r)=\frac{1}{N_{\zeta_{i}}} e^{-\zeta_{i} r}$ 
and $\int d\vec r \rho_{i}(\vec r)=1$. The 
exponent $\zeta_{i}$ is chosen 
such that the electron-electron repulsion energy due to two such charge 
distributions on atomic-site $i$ equals the difference between the 
atomic electronic affinity and ionization potential 
$\int d\vec r d\vec r' \rho_{i}(\vec r) \rho_{i}(\vec r')/|\vec r- \vec r'|  
=I_{i}-A_{i}$~\cite{Frauenheim}, which incorporates implicitly the 
short-range on-site electron-electron interaction effect. In this way, 
the change in the electrostatic potential can be written as superposition 
of atomic-centered potentials 
$\delta V(\vec r)=\sum_{i} \delta N_{i} V_{i}(\vec r -\vec r_{i})$. The 
advantage of the present approximation is that 
$V_{i}(\vec r -\vec r_{i})
=\int d\vec r' \rho_{i}(\vec r'-\vec r_{i})/|\vec r- \vec r'|$ 
can be evaluated analytically,~\cite{Frauenheim,XueMRS03} 
\begin{equation}
\label{Vi}
V_{i}=(1-e^{-\zeta_{i}|\vec r-\vec r_{i}|}
(1+\zeta_{i}|\vec r-\vec r_{i}|/2))/|\vec r-\vec r_{i}|.   
\end{equation}

For the metal-SWNT-metal junction considered here, we take into account the 
image-potential effect by including within $\delta V$ contributions from 
both atom-centered charges and their image charges (centered around the 
image positions), rather than imposing an image-type potential correction on  
$\delta V$. The charge transfer-induced electrostatic potential change is thus: 
\begin{equation}
\label{VES}
\delta V(\vec r)=\sum_{i} [\delta N_{i} V_{i}(\vec r -\vec r_{i}) 
 + \delta N_{i;image} V_{i}(\vec r -\vec r_{i;image})]
\end{equation}
where the image charges $\delta N_{i;image}$ and their positions 
$\vec r_{i;image}$ are determined from standard electrostatics 
considerations.~\cite{CED,NoteImag} The self-consistent cycle proceeds 
by evaluating the matrix elements of the potential 
$\delta V_{mn}=\int d\vec r \phi_{m}^{*}(\vec r) \delta V(\vec r) 
\phi_{n}(\vec r)$ using two types of scheme: (1) If $m,n$ belong to 
the same atomic site $i$, we calculate it by direct numerical integration; 
(2) If $m,n$ belong to different atomic sites, we calculate it from the 
corresponding on-site element using the approximation 
$\delta V_{mn}=1/2S_{mn}(\delta V_{mm}+\delta V_{nn})$ where 
$S_{mn}$ is the corresponding overlap matrix element. We also calculate 
the matrix elements of the external potential $V_{ext}$ by direct numerical 
integration whenever applicable. Given the Hamiltonian matrix 
$H=H_{0}+V_{ext}+\delta V$, the self-consistent calulation then proceeds 
by calculating the density matrix $\rho$ from 
the Green's function by integrating over a complex energy 
contour~\cite{Xue02,XueMol1,XueMol2,Contour} and 
evaluating the net charge on 
atomic-site $i$ from $\delta N_{i}=(\rho S)_{ii}-N_{i}^{0}$ where 
$N_{i}^{0}$ is the number of valence electrons on atomic-site $i$ 
of the bare SWNT. Note that the 
advantage of the present self-consistent tight-binding treatment 
is that \emph{no adjustable parameters have been introduced} besides 
those that may be present in the initial Hamiltonian $H_{0}$. 

Once the self-consistent calculation converges, we can calculate the 
transmission coefficient through the SWNT junction from
\begin{equation}
\label{TEV}
T(E,V)=
Tr[\Gamma_{L}(E,V)G^{r}(E,V)\Gamma_{R}(E,V)[G^{r}]^{\dagger}(E,V)],
\end{equation}
and the spatially-resolved local density of states (LDOS) from 
\begin{equation}
\label{SLDOS}
n(\vec r,E)=-\frac{1}{\pi} \lim_{\delta \to 0^{+}} \sum_{ij}
Imag[G_{ij}^{r}(E+i\delta)] \phi_{i}(\vec r) \phi_{j}^{*}(\vec r), 
\end{equation}
The spatial integration of LDOS gives the density of states,
\begin{equation}
\label{DOS}
n^{\sigma}(E)=\int d\vec r n^{\sigma}(\vec r,E)
= -\frac{1}{\pi} \lim_{\delta \to 0^{+}} 
  Tr\{Imag[G^{r}(E+i\delta)] S\}=\sum_{i}n_{i}(E) 
\end{equation}
where the atomic site-resolved density of states is 
$n_{i}(E)= -\frac{1}{\pi} \lim_{\delta \to 0^{+}} 
[Imag[G^{r}(E+i\delta)] S]_{ii}$. 
Within the coherent transport regime, the terminal current is related to the 
transmission coefficient through the 
Landauer formula~\cite{MesoPhy,Buttiker86,LB88}
\begin{equation}
\label{IV}
I=\frac{2e}{h} \int dE T(E,V) [ f(E-\mu_{L})-f(E-\mu_{R}) ]
\end{equation}
where we can separate the current into two components, the 
``tunneling'' component $I_{tun}$ and the ``thermionic emission'' 
component $I_{th}$ as follows,
\begin{equation}
I = I_{tun}+I_{th} \nonumber 
  = \frac{e}{h} [\int_{\mu_{L}}^{\mu_{R}}+
  (\int_{-\infty }^{\mu_{L}}+\int^{+\infty }_{\mu_{R}})] 
 dE T(E,V)[f(E-\mu_{L})-f(E-\mu_{R})] 
\end{equation}
Similarly, we can separate the zero-bias conductance 
\begin{equation}
\label{GT}
G=\frac{2e^{2}}{h}\int dE T(E)[-\frac{df}{dE}(E-E_{F})]=G_{Tu}+G_{Th}
\end{equation}
into the tunneling contribution 
$G_{tun}=\frac{2e^{2}}{h}T(E_{F})$ and thermal-activation 
contribution $G_{th}=G-G_{tun}$. 

\subsection{Device model}

In this work, we take $(10,0)$ SWNT as the protype semiconducting SWNT. 
Since for metal-semiconductor contacts, high- and low- work function metals 
are used for electron injection and hole-injection respectively, we consider 
both gold (Au) and titanium (Ti) electrodes as examples of high- and 
low- work function metals (with work functions of $5.1$ and $4.33$ eV 
respectively~\cite{Note,CRC}). The work function of the (10,0) SWNT is taken 
the same as that of the graphite ($4.5$ eV)~\cite{DekkerNT,AvNT}. 
In this paper, the Hamiltonian $H_{0}$ describing the bare SWNT is 
obtained using the semi-empirical Extended Huckel Theory (EHT) with 
corresponding non-orthogonal Slater-type basis sets 
$\phi_{m}(\vec r)$~\cite{Hoffmann88,AvEHT} describing the valence ($sp$)  
electrons of carbon, while the self-energy due to the contact to the metallic 
electrodes is evaluated using tight-binding parameters obtained from fitting 
accurate bulk band structure.~\cite{Xue02,Papa86} The calculation is 
performed at room temperature. 

The $(10,0)$ SWNT has a diameter of $7.8(\AA)$ and unit cell length of 
$4.1(\AA)$. The unit cell consists of 4 carbon rings with 10 carbon atoms 
each. The calculated bulk band gap using EHT is $\approx 0.9(eV)$. 
Since the contacts involved in most transport measurement are not well 
characterized, a microscopic study as presented here necessarily requires 
a simplified model of the interface, which is illustrated schematically in 
Fig.\ \ref{xueFig1}.  Here the finite SWNT molecules are 
attached to the electrode surface through the ring of dangling-bond 
carbon atoms at the ends. We neglect the possible distortion of the SWNT 
atomic structure induced by the open-end and its subsequent adsorption 
onto the electrode surface.~\cite{Contact} We assume that the axis of 
the SWNT molecule (the Z-axis) lies perpendicular to the electrode surface 
(the XY-plane). Only nearest-neighbor metal atoms on the surface layer of the 
electrode are coupled to the SWNT end, the surface Green's functions of 
which are calculated using the tight-binding parameter of Ref.\ 
\onlinecite{Papa86} assuming a semi-inifinte substrate corresponding 
to the $\langle 111 \rangle$ and hcp surface for the gold and titanium 
electrodes respectively.  

\begin{figure}
\includegraphics[height=3.2in,width=4.0in]{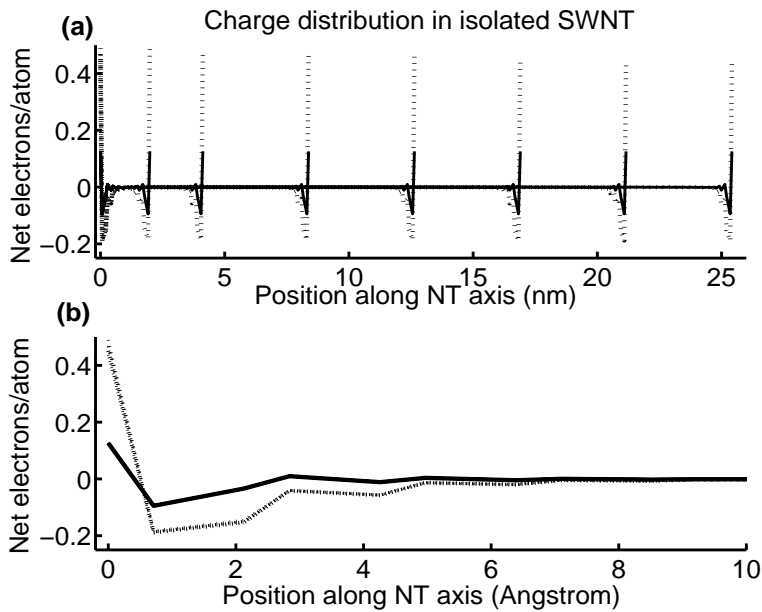}
\caption{\label{xueFig2} (a) shows net electron distribution in 
the isolated SWNT molecule as a function of SWNT length for seven 
different lengths. (b) shows magnified view at the left end of all 
SWNT molecules studied. Here the solid lines show the results 
obtained using EHT with self-consistent correction, while the dotted lines 
show the results obtained using EHT without self-consistent correction.  }
\end{figure}

\begin{figure}
\includegraphics[height=3.2in,width=4.0in]{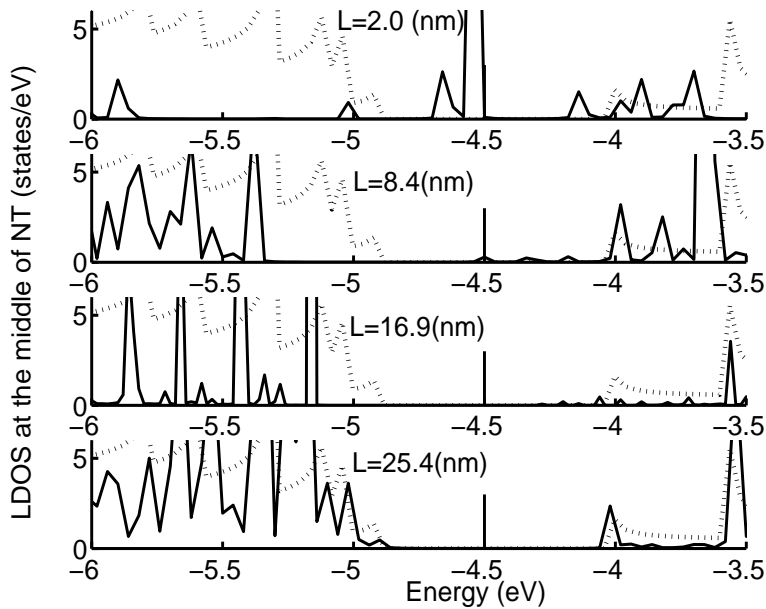}
\caption{\label{xueFig3}
Local density of states in the middle of the $(10,0)$ SWNT molecule 
calculated using the self-consistent EHT for SWNT lengths of 
$2.0,8.4,16.9$ and $25.4(nm)$ respectively. The vertical line at 
$E=-4.5(eV)$ denotes the Fermi-level position of the bulk SWNT.  
The dotted line is the LDOS of the bulk $(10,0)$ SWNT. For clarity, the 
figures have been cut off at the top where necessary.  } 
\end{figure}

The lengths of the $(10,0)$ SWNT molecule investigated are $L=2.0,4.1,
8.4,12.6,16.9,21.2$ and $25.4$ (nm), corresponding to 
$5,10,20,30,40,50$ and $60$ unit cells respectively. As discussed in the 
following sections, the variation of SWNT length from $5$ to $60$ unit 
cells spans the entire range from the molecular limit to the 
bulk limit. To evaluate the dependence of Schottky barrier formation on the 
strength of metal-SWNT interface coupling, we consider three SWNT 
end-metal surface distances of $\Delta L = 2.0,2.5$ and $3.0 (\AA)$. 
Note that the average of the nearest-neighbor atom distance in the 
SWNT and Au/Ti electrode is around $2.1 (\AA)$. From our previous work on 
first-principles based modeling of molecular electronic devices,~\cite{XueMol2} 
we find that increasing metal-molecule distance by $1.0(\AA)$ is sufficient 
to reach the weak interfacial coupling limit. Therefore the three choices of 
metal-SWNT distance are sufficient to demonstrate the trend of Schottky 
barrier formation as the strength of interface coupling varies from the strong 
coupling to the weak coupling limit.  
 
\section{Evolution of the electronic structure of the SWNT molecule 
with length}

\begin{figure}
\includegraphics[height=3.2in,width=5.0in]{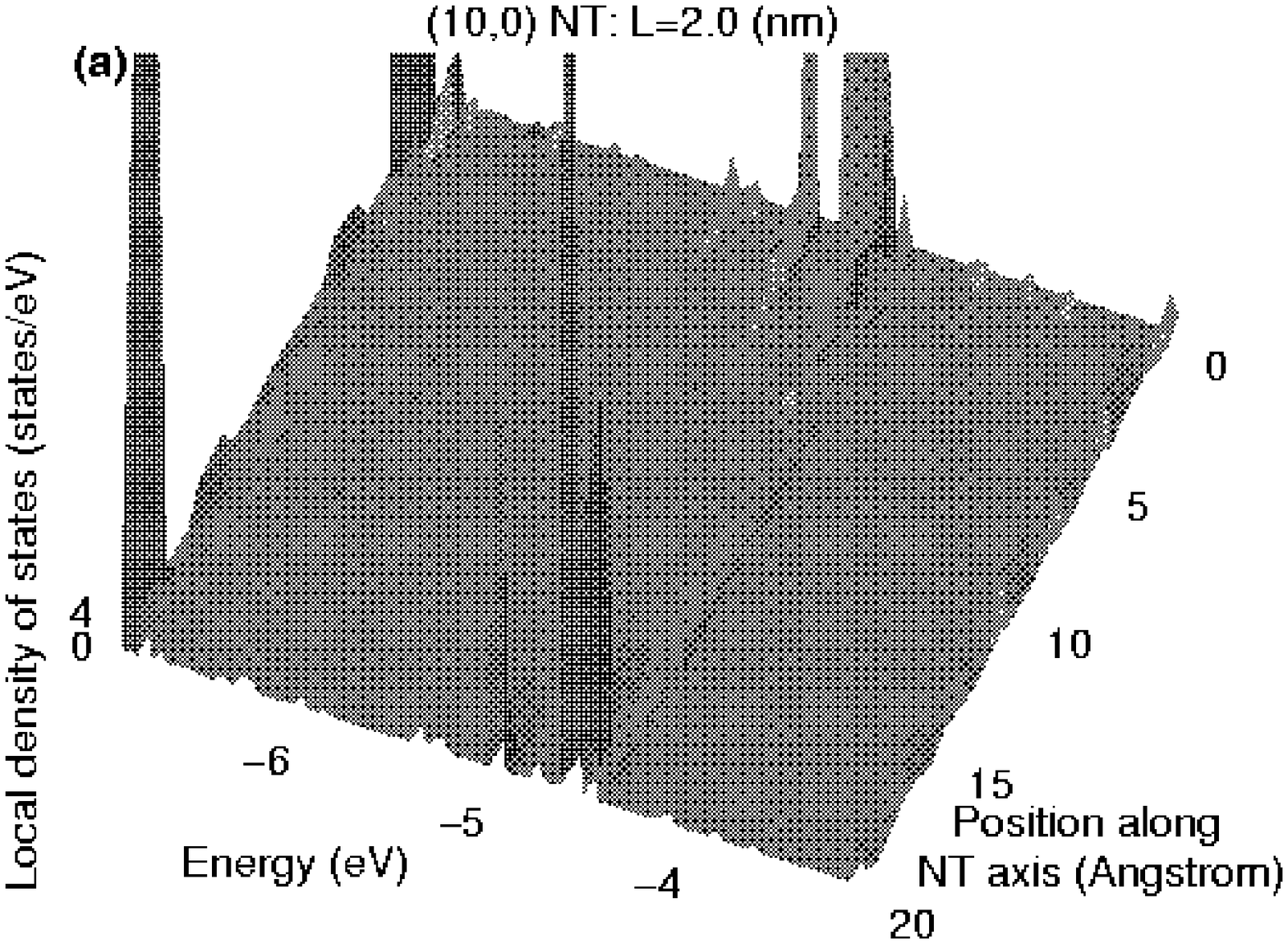}
\includegraphics[height=3.2in,width=5.0in]{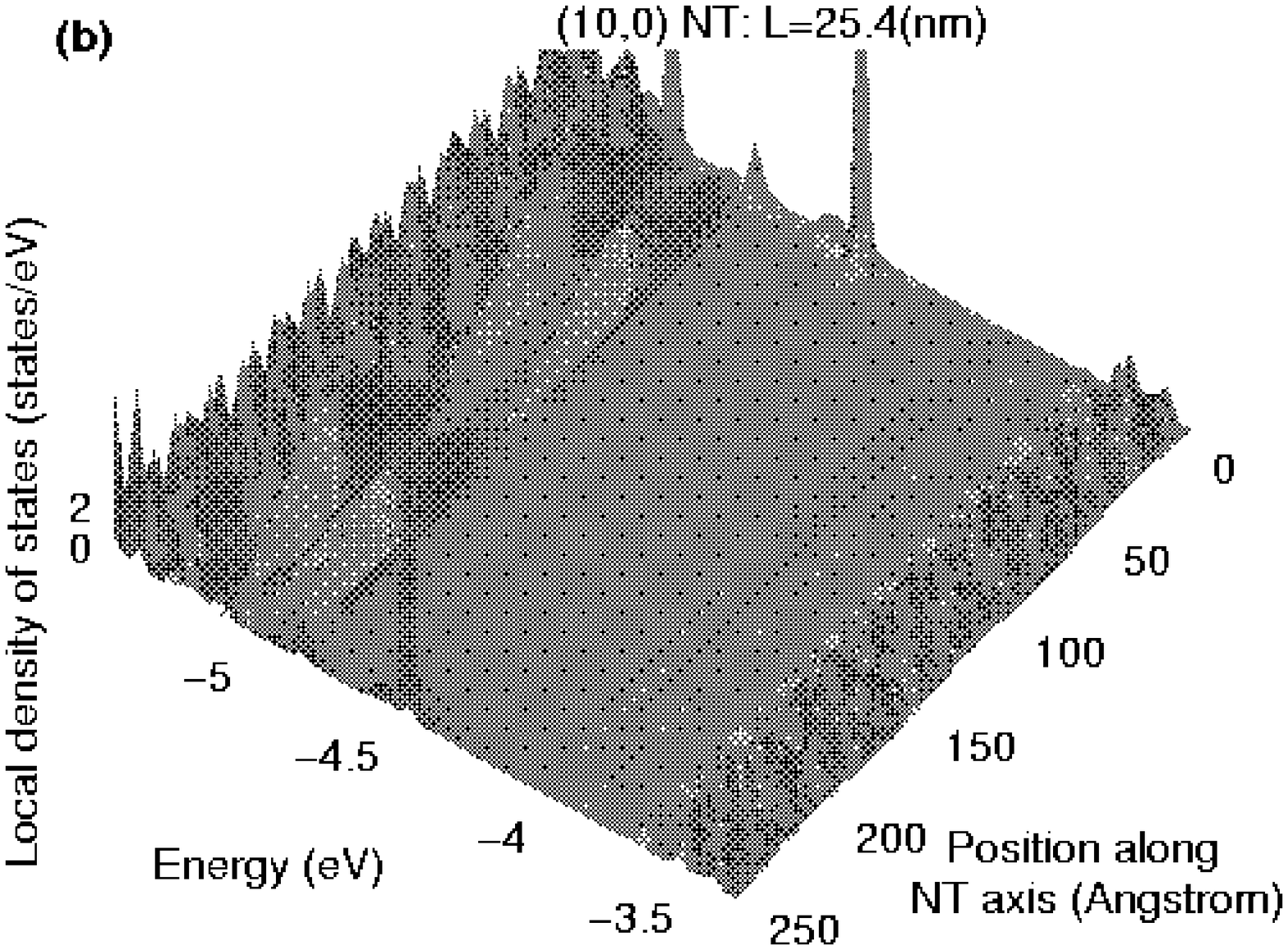}
\caption{\label{xueFig4}  (Color online)
Local density of states as a function of position along the axis of the 
$5$-unit cell (a) and $60$-unit cell (b) SWNT molecule. The LDOS is 
obtained by summing over the 10 carbon atoms of each ring 
of the $(10,0)$ SWNT. Each cut along the energy axis for a given position 
along the NT axis gives the LDOS at the corresponding carbon ring. 
Each cut along the postion axis for a given energy illustrates the spatial 
extension of the corresponding electron state. For the $5$-unit cell 
SWNT(a), localized dangling bond state exists around $-5.0(eV)$, whose 
wavefunction decays into the interior of the SWNT molecule. For the 
$60$-unit cell SWNT which has approached the bulk limit, the localized 
dangling bond state is located instead around $-4.5(eV)$, .i.e., the 
middle of the conduction/valence band gap.  } 
\end{figure}

The dangling $\sigma$ bonds at the open end of the SWNT molecule 
lead to charge transfer 
between carbon atoms at the end and carbon atoms in the interior 
of the SWNT. This should be corrected self-consistently first and gives 
the intial charge configuration $N_{i}^{0}$for determining the 
charge transfer within the metal-SWNT-metal junction in later sections.  
The self-consistent calculation proceeds as described in the previous section, 
except that there is no self-energy operator associated with the contact in 
the case of the bare SWNT molecule. The result is shown in 
Fig.\ \ref{xueFig2}, where we plot the net electrons per atom as a 
function of position along the $(10,0)$ SWNT axis obtained from 
both EHT and the self-consistent EHT calculations. The self-consistent 
treatment suppresses both the magnitude and the range 
of the charge transfer, which are approximately the same for all the 
SWNT molecules investigated, reflecting the localized nature of the 
perturbation induced by the end dangling bonds (Fig\ \ref{xueFig2}(b)). 

To evaluate the evolution of SWNT electronic structure with molecule 
length, we calculated the local density of states in the middle unit cell of the 
SWNT molecule using the self-consistent EHT and compared them with those 
of the bulk (infinitely long) SWNT. The results for SWNT lengths of 
$2.0,8.4,16.9$ and $25.4(nm)$ are shown in Fig.\ \ref{xueFig3}. 
Here the LDOS of the isolated finite SWNT molecule is artificially broadened 
by inserting a small but finite imaginary number ($\delta =10^{-6} (eV)$) 
into the retarded Green's function $G^{r}(E+i\delta )$. Therefore only the 
band edge location but not the exact value of the LDOS should be examined 
when evaluating the approach to the bulk limit with increasing nanotube 
length. The LDOS of the shortest SWNT molecule ($2.0$ nm) shows completely 
different structure from that of the bulk. In particular, there are peaks located 
within the conduction-valence band gap of the bulk SWNT caused by the 
localized dangling bond states at the end, which decays into the 
interior of the short SWNT molecule. This is illustrated by the 
position-dependent LDOS along the NT axis in Fig.\ \ref{xueFig4}. 
We can therefore characterize the $5$-unit cell SWNT as being in 
the molecular limit. The magnitude of the localized dangling-bond states 
in the middle is suppressed exponentially with increasing SWNT length 
and is negligible for all other SWNT molecules studied. 

The development of the SWNT valence bands with molecule length is 
clear from Fig.\ \ref{xueFig3}. The development of the SWNT conduction 
bands is less regular since tight-binding theory constructed for valence 
electrons generally describes the valence bands better than the conduction 
bands.~\cite{Chadi75} The approach to the bulk band structure is 
obvious for SWNT longer than $40$ unit cells and complete for the SWNT 
molecules of $60$ unit cells long. Therefore the variation of SWNT 
length from $5$ to $60$ 
unit cells spans the entire range from the molecular limit to the bulk limit. 
Note that as the length of the SWNT molecule changes, the energy of 
localized dangling bond states also changes, which saturates as the SWNT 
approaches the bulk limit. For the $60$-unit cell SWNT, it is located around 
$-4.5(eV)$, i.e., the Fermi-level of the bulk SWNT. This is consistent with the 
previous observation in semiconductor interfaces, where it has been argued 
that the dangling bond level plays the role of ``charge-neutrality-level'' 
(CNL) in band lineup involving semiconductors, which is located around 
the midgap for semiconductors with approximately symmetric conduction 
and valence band structures.~\cite{CNL,TersoffMS} 

\section{Scaling analysis of Schottky barrier formation at metal-SWNT 
molecule interfaces}

\subsection{Schottky barrier formation at planar metal-semiconductor 
interfaces}

\begin{figure}
\includegraphics[height=3.2in,width=4.5in]{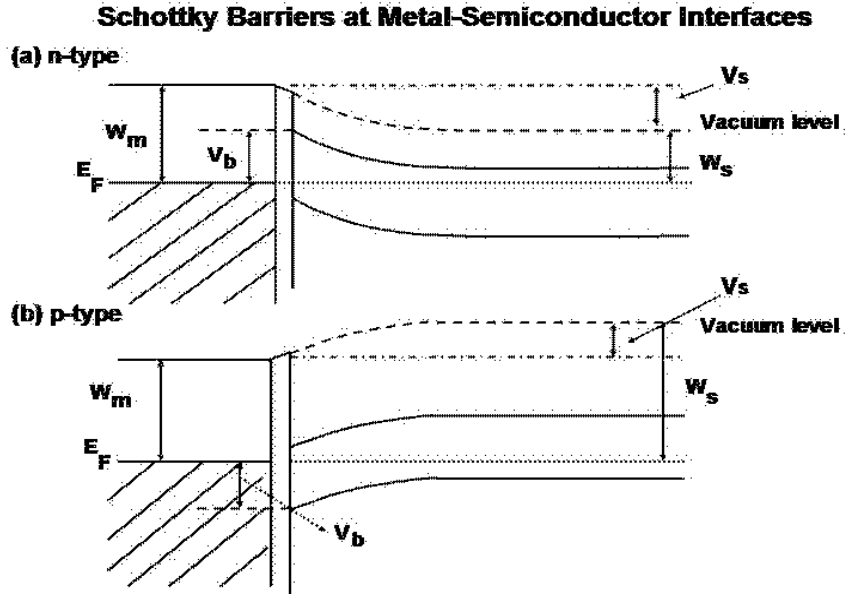}
\caption{\label{xueFig5}
Schematic illustration of the formation of Schottky barrier at the planar  
metal-semiconductor interfaces. (a) n-type semiconductor; 
(b) p-type semiconductor. $W_{m},W_{s}$ are the work functions of 
the metal and semiconductor respectively. $V_{b}$ is the Schottky 
barrier height for electron (hole) injection at the n-type (p-type) 
semiconductor interface. $V_{s}$ is the additional potential shift inside 
the semiconductor due to the depleted dopant charges.  }
\end{figure}

We start with a brief summary of Schottky barrier formation at an ideal 
planar metal-semiconductor interface~\cite{MSMonch,MSReview,TersoffMS} 
to motivate our discussion of metal-SWNT interface in later sections. 
An ideal metal-semiconductor interface is formed by reducing the 
distance between a metal and a semi-infinite semiconductor until 
an intimate and abrupt interface forms,~\cite{MSMonch} as 
illustrated in Fig. \ref{xueFig5}.

The open-end of the semi-infinite semiconductor leads 
to localized surface states whose wavefunctions decay exponentially 
into the vacuum and inside the semiconductor, the nature of which  
can be undersood qualitatively from the complex band structure of the 
bulk semiconductor by extrapolating the energy band 
into the band gap region. Upon contact with the metal electrodes, the 
intrinsic semiconductor surface states are replaced by Metal-Induced 
Gap States (MIGS), which are the tails of the metal wavefunction 
decaying into the semiconductor within the band gap since the 
wavefunctions there are now matched to the continuum of states 
around the metal Fermi-level.~\cite{BH,Louie} The corresponding charge 
transfer induces an interface dipole layer due to the planar structure, the 
electrostatic potential drop across which shifts rigidly the 
semiconductor band relative to the metal Fermi-level $E_{F}$. 
Additional electrostatic potential 
change can also occur if the semiconductor is doped and a space-charge 
layer forms due to the depleted dopant charges, as illustrated in 
Fig.\ \ref{xueFig5}. The total potential shift must be such that the two 
Fermi-levels across the interface line up.  The potential variations 
away from the interface dipole layer introduced by the space 
charge layer are slow (on the order of magnitude of $\sim 0.5 (V)$ 
within hundreds of nm or longer) due to the small percentage of 
dopant atoms.~\cite{MSBook} This leads to the picture of band shift 
following electrostatic potential 
change since such potential variation occurs on a length scale much longer 
than the semiconductor unit cell size. 

The band lineup at the planar metal-semiconductor interface is  
determined by the overall charge neutrality condition and the 
corresponding one-dimensional electrostatic considerations: 
$Q_{m}+Q_{is}+Q_{sc}=0$,
where $Q_{m}$, $Q_{is}$ and $Q_{sc}$ are the surface charge density 
within the metal (m) surface layer, semiconductor surface layer due to the 
interface states (is) and semiconductor space-charge (sc) layer respectively, 
which are obtained by averaging the three-dimensional charge density over 
the plane parallel to the interface. For n(p)-type semiconductor, 
the Schottky barrier height $V_{b}$ for 
electron (hole) injection is determined by $E_{F}$ and the conduction 
(valence) band edge. Since electrons can easily tunnel through the 
interface dipole layer, current tranport occurs by charge carriers injected 
into the bulk conduction/valence band states by tunneling through or 
thermionically emitted over the interface barrier. So the Schottky barrier 
height alone can be used for characterizing the transport 
characteristics.~\cite{MSBook} 

Two key concepts thus underlie the analysis of Schottky barrier 
formation at the planar metal-semiconductor interface: (1) The separation 
into the interface region (dipole layer) and the bulk semiconductor region 
(including the space-charge layer) with well-defined Fermi-level; (2) The 
rigid band shift following the local electrostatic potential change due to 
the planar interface structure. Both concepts are not valid in analyzing 
Schottky barrier formation at metal-SWNT interfaces.   
    
\subsection{Electrostatics of the metal-SWNT molecule interface}

\begin{figure}
\includegraphics[height=3.2in,width=4.0in]{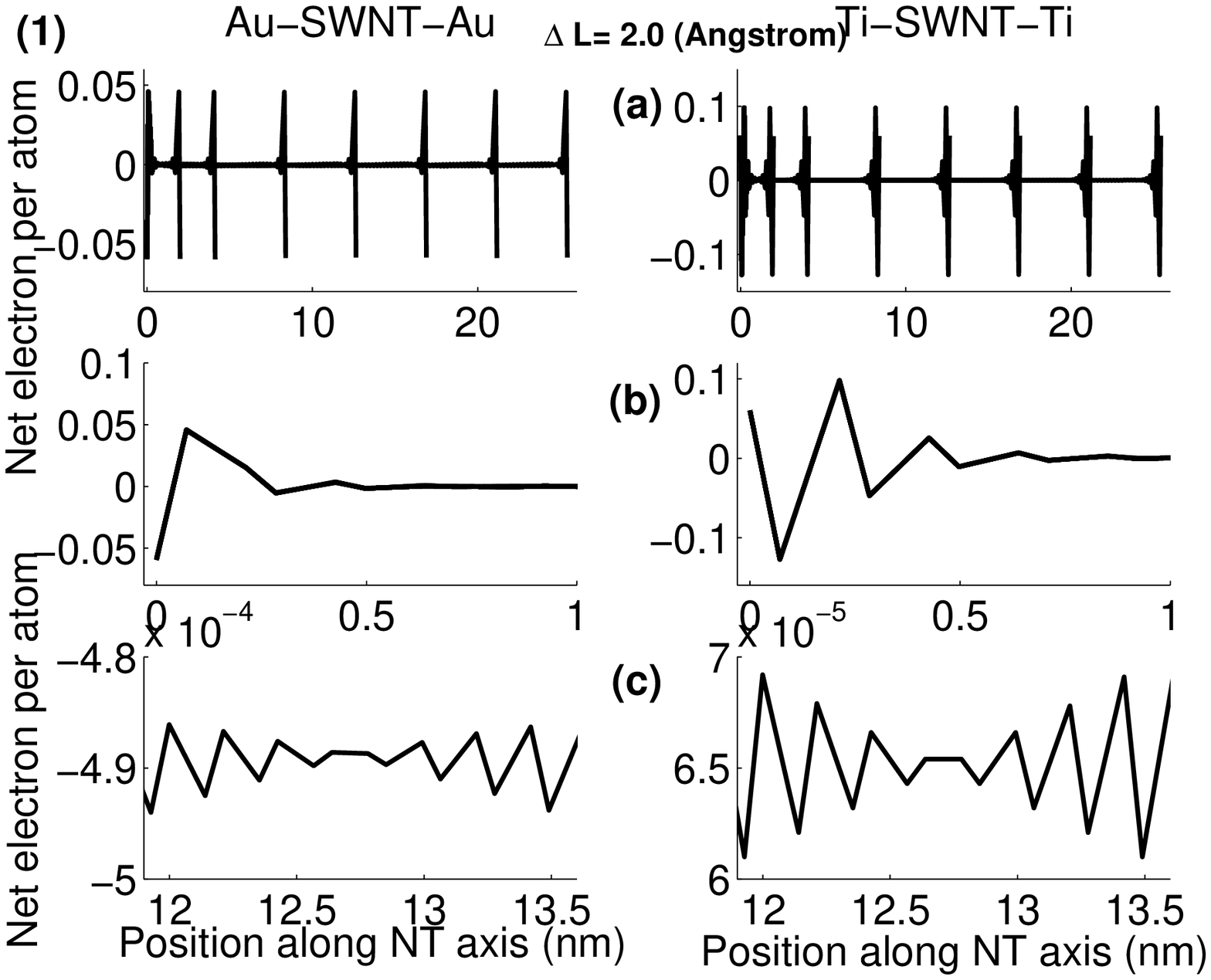}
\includegraphics[height=3.2in,width=4.0in]{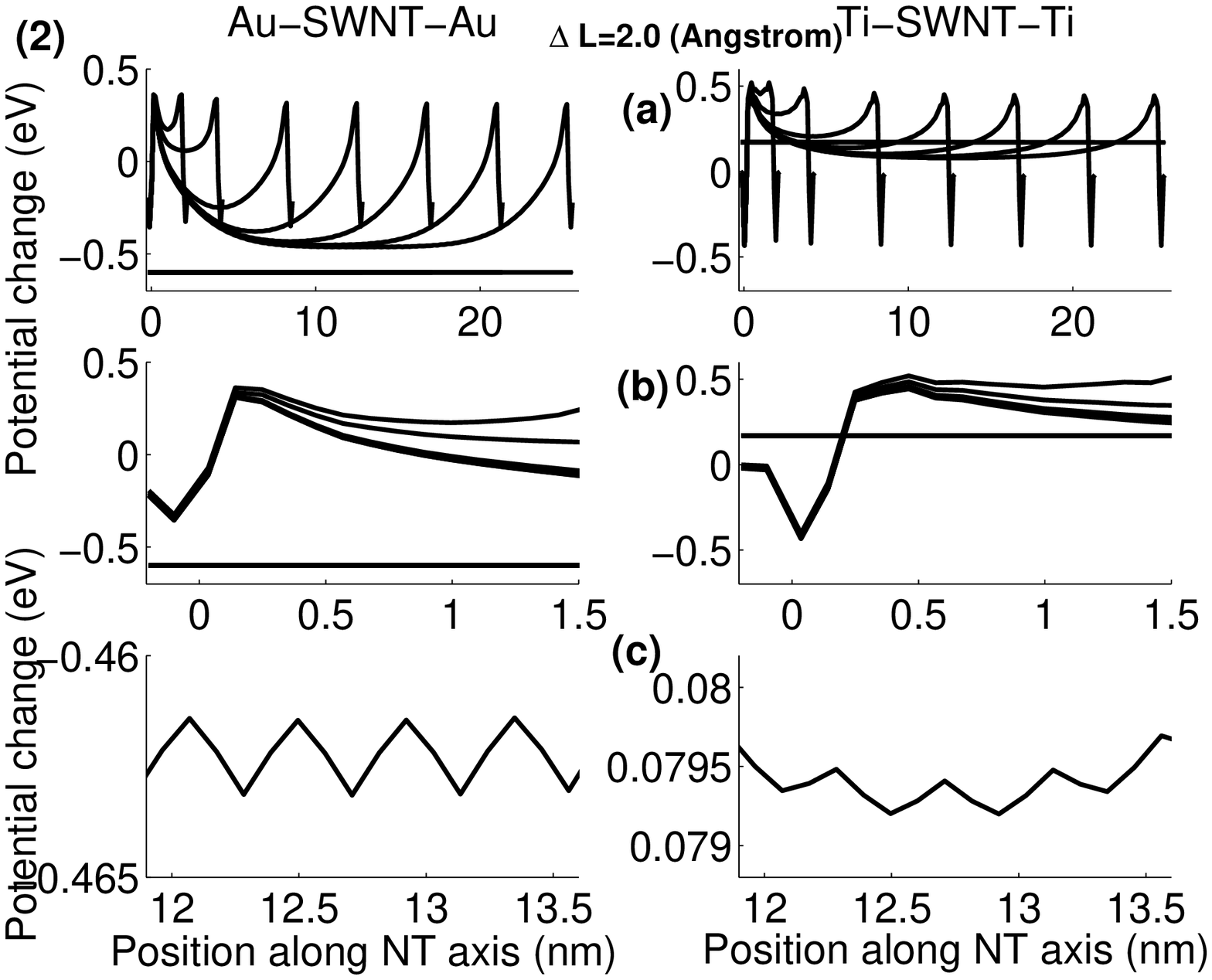}
\caption{\label{xueFig6} 
Charge transfer (1) and electrostatic potential change (2) at the 
Au-finite SWNT-Au and Ti-finite SWNT-Ti junctions as a function of 
SWNT length for seven different lengths at SWNT-metal distance 
of $\Delta L=2.0(\AA)$. For each junction, we have also shown the 
magnified view both at the metal-SWNT interface (b) and in the middle 
of the longest (25.4 nm) SWNT molecule (c).  The horizontal lines in the 
potential plot (2) denote the work function differences 
between the electrodes and the bulk SWNT.  }
\end{figure}

\begin{figure}
\includegraphics[height=3.2in,width=4.0in]{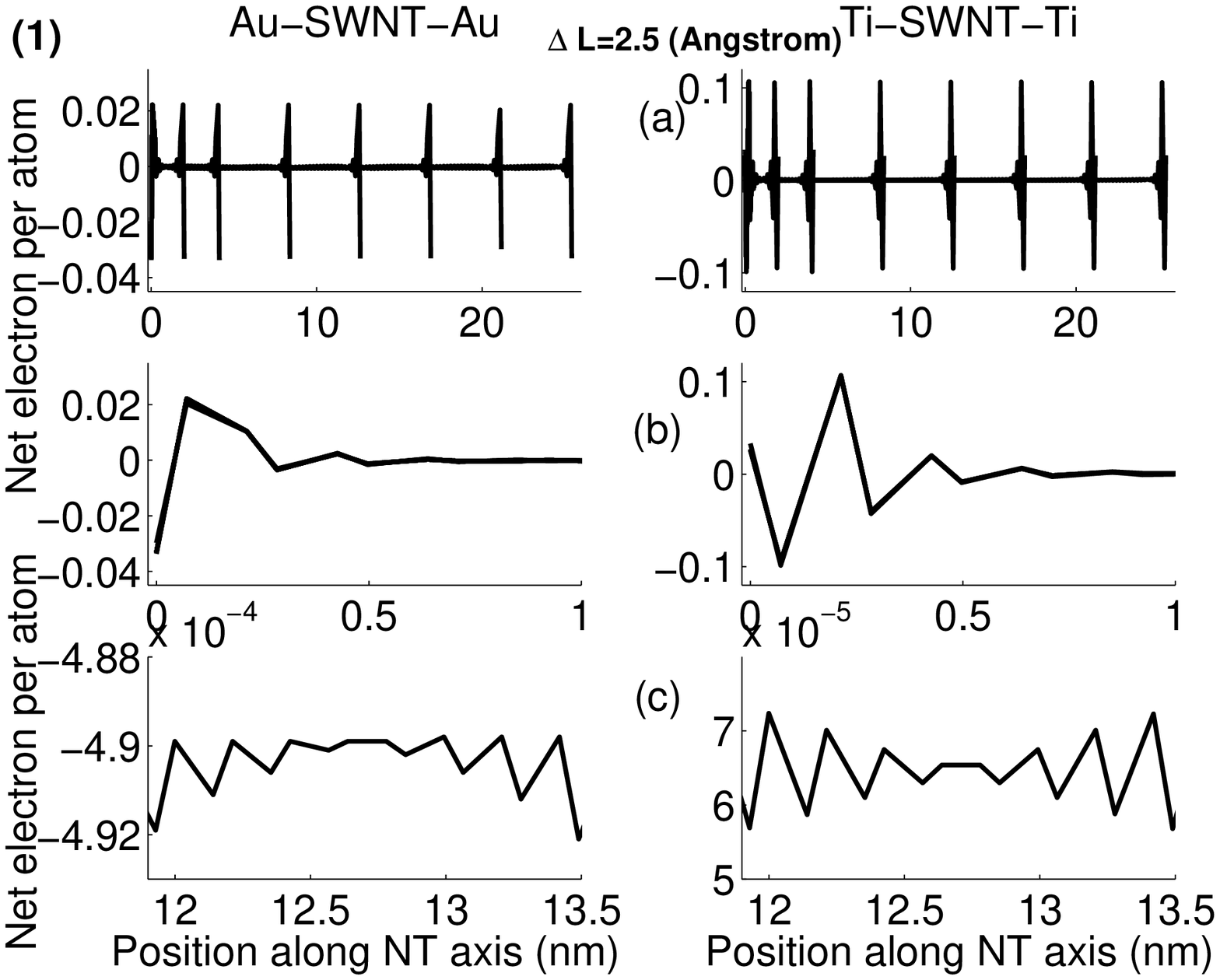}
\includegraphics[height=3.2in,width=4.0in]{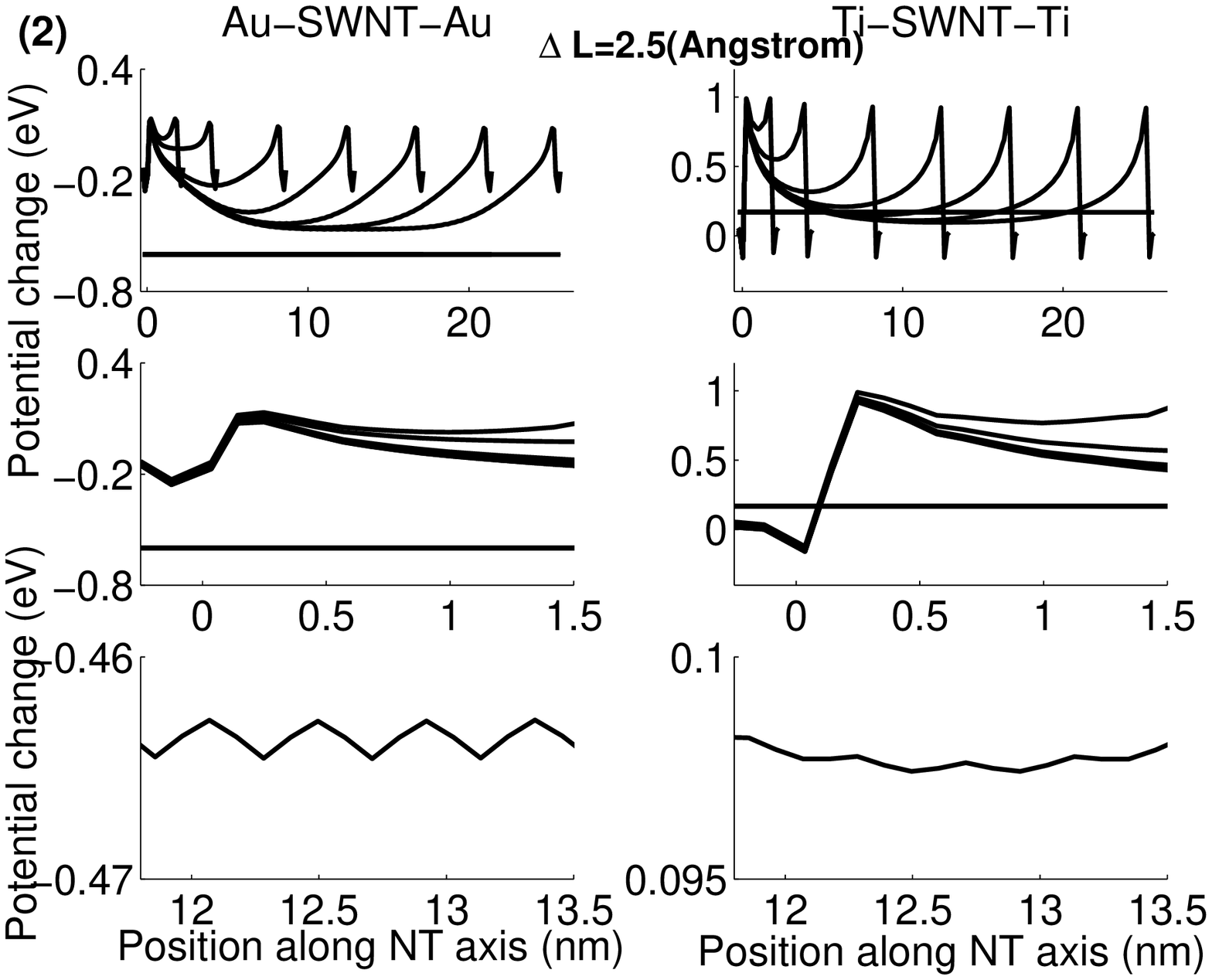}
\caption{\label{xueFig7} 
Charge transfer (1) and electrostatic potential change (2) at the 
Au-finite SWNT-Au and Ti-finite SWNT-Ti junctions as a function of 
SWNT length for seven different lengths at SWNT-metal distance 
of $\Delta L=2.5(\AA)$. For each junction, we have also shown the 
magnified view both at the metal-SWNT interface (b) and in the middle 
of the longest (25.4 nm) SWNT molecule (c). The horizontal lines in the 
potential plot (2) denote the work function differences 
between the electrodes and the bulk SWNT.   }
\end{figure}

\begin{figure}
\includegraphics[height=3.2in,width=4.0in]{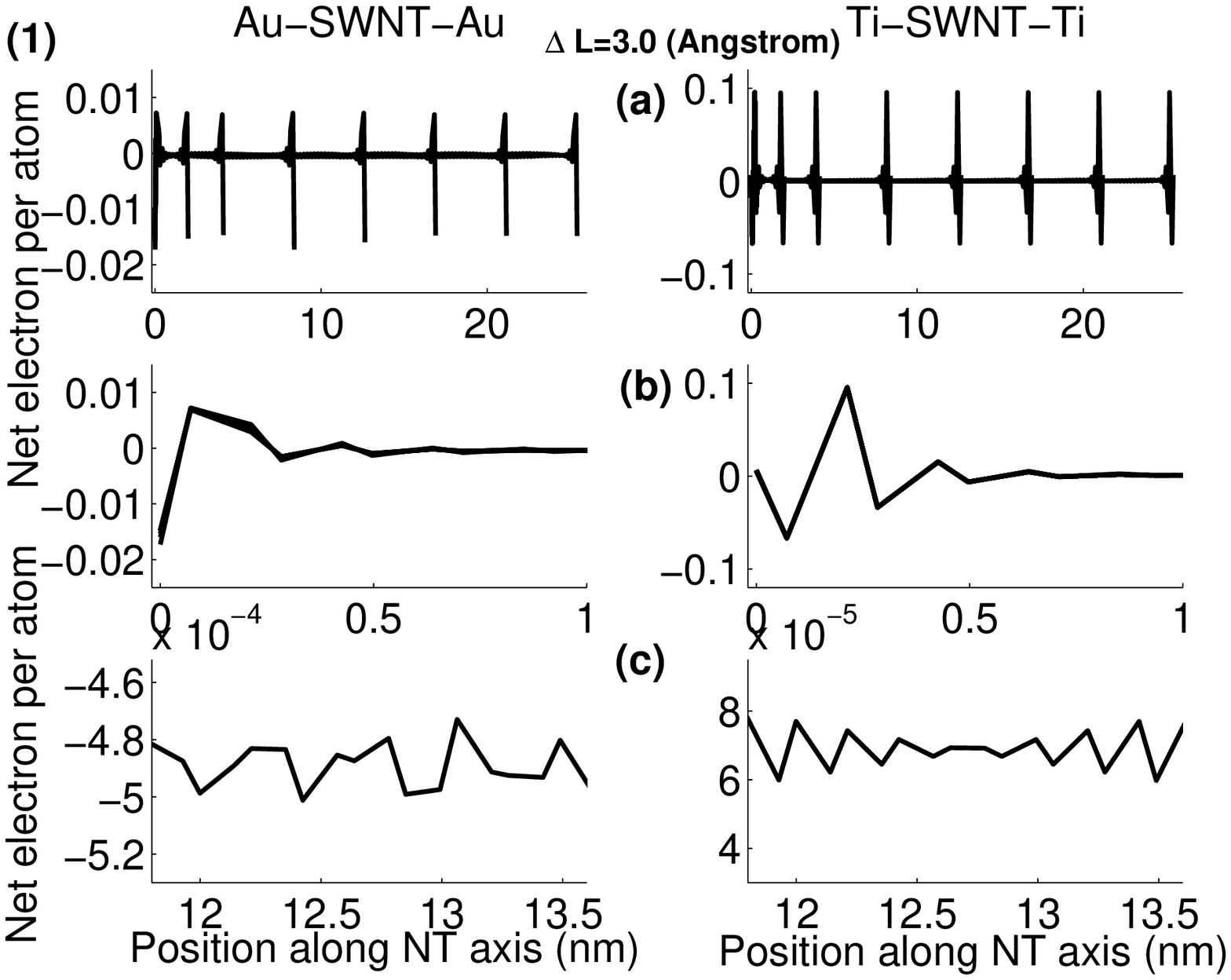}
\includegraphics[height=3.2in,width=4.0in]{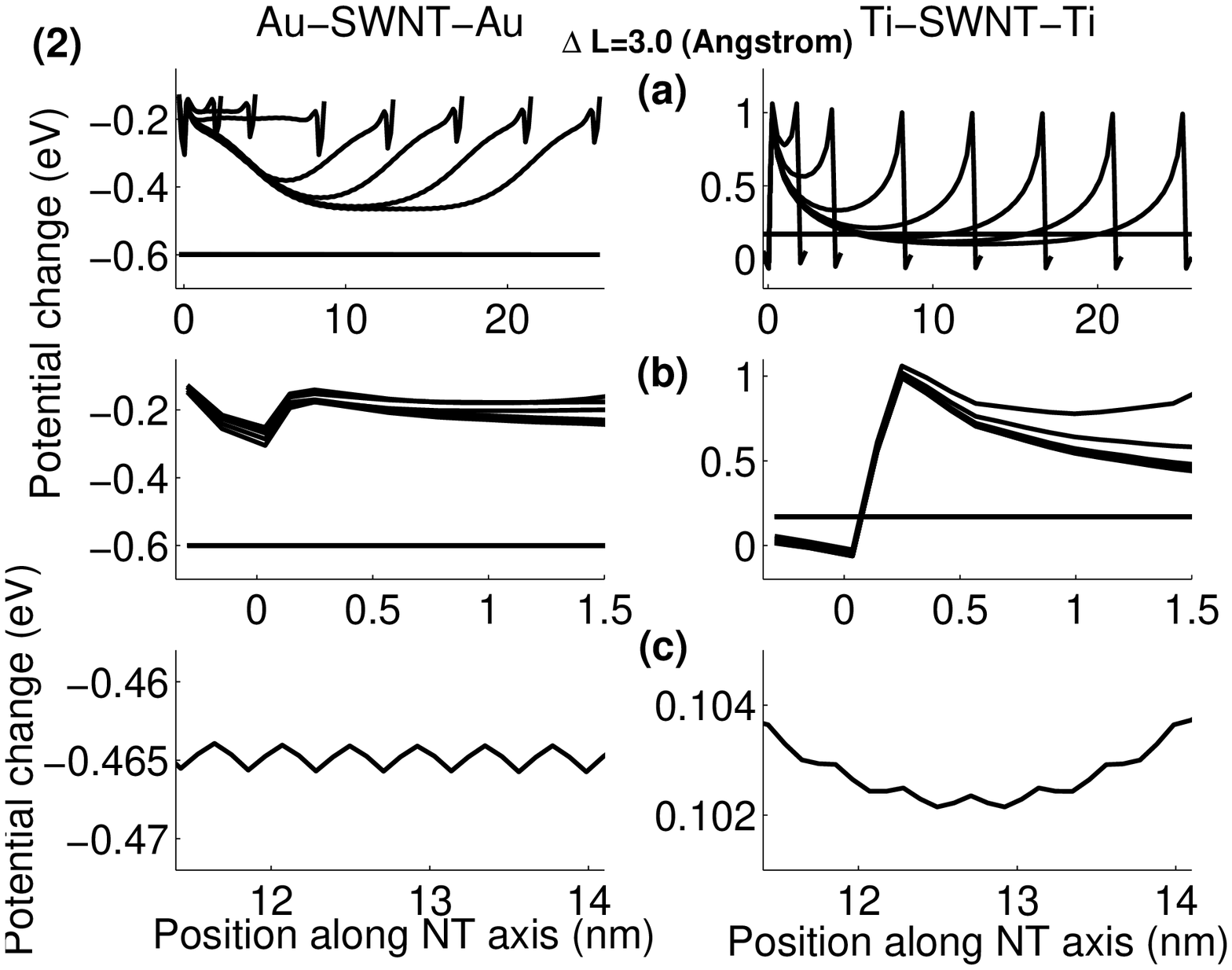}
\caption{\label{xueFig8} 
Charge transfer (1) and electrostatic potential change (2) at the 
Au-finite SWNT-Au and Ti-finite SWNT-Ti junctions as a function of 
SWNT length for seven different lengths at SWNT-metal distance 
of $\Delta L=3.0(\AA)$. For each junction, we have also shown the 
magnified view both at the metal-SWNT interface (b) and in the middle 
of the longest (25.4 nm) SWNT molecule (c). The horizontal lines in the 
potential plot (2) denote the work function differences 
between the electrodes and the bulk SWNT.  }
\end{figure}

The calculated charge transfer and electrostatic potential change at 
the gold-SWNT-gold and titanium-SWNT-titanium junctions are shown in 
Figs. \ref{xueFig6}-\ref{xueFig8} for metal-SWNT distance of 
$\Delta L=2.0,2.5,3.0(\AA)$ respectively. The electrostatic potential 
change is obtained as the difference between electrostatic potentials 
within the metal-SWNT-metal junction and the bare SWNT molecule, 
which is calculated from the transferred charge throughout the SWNT 
using Eq. \ref{VES}. Due to the molecular-scale dimension of both the 
SWNTs and the contact area, the transferred charge across the interface 
is confined in a finite region. Unlike the dipole \emph{layer} at the bulk 
metal-semiconductor interface which induces a step-wise change in the 
electrostatic potential, the transferred charge across the 
metal-SWNT interface takes the form of molecular-size dipole, the 
electrostatic potential of which \emph{decays to zero in 
regions far away from the interface}.~\cite{TersoffNT,Odin00} 
In addition, the SWNT molecule is undoped. The occupation of the 
electron states within the SWNT is determined by the Fermi-level of 
the electrodes, even for a long SWNT which has reached the bulk limit 
and a Fermi-level can be defined from the bulk band structure.   

Note that despite the delocalized nature of SWNT electron states 
in the conduction/valence band, for a given metal-SWNT 
distance $\Delta L$, both the magnitude and the range of the charge 
transfer at the metal-SWNT molecule interface are approximately 
independent of the SWNT length, reflecting the localized nature of 
interfacial charge transfer process~\cite{XueMol1,XueMol2} The charge 
transfer adjacent to the metal-SWNT interface shows Friedel-like 
oscillation.~\cite{Friedel} Such Friedel-like oscillations of transferred 
charge have also been observed in planar metal-semiconductor 
interfaces,~\cite{Louie} finite atomic chains~\cite{LangAv} and 
molecular tunnel junctions.~\cite{XueMol1,XueMol2} 
The oscillation of the interface-induced charge transfer dies out quickly 
inside the SWNTs as the length of the SWNT molecule increases. 
The oscillation in both the transferred charge and electrostatic potential 
change in the middle of the SWNT are due to the intrinsic 
two-sublattice structure of the zigzag tube, and persist in an infinitely long 
zigzag tube.~\cite{XueNT04,TersoffNT02} 

As $\Delta L$ increases from $2.0 \AA$ to $3.0 \AA$, the magnitude of the 
charge transfer oscillation at the 
interface decreases with the decreasing interface coupling strength, but 
the magnitude of charge transfer inside the SWNT molecule is almost 
independent of the coupling strength across the interface. For the 
Au-SWNT-Au junction, there is a small positive charge transfer of 
$4.9 \times 10^{-4}$ per atom in the middle of the $60$-unitcell SWNT, 
while for the Ti-SWNT-Ti junction, there is instead a small negative 
charge tranfer of $-6.5\times 10^{-5}$ per atom.~\cite{Note2}   

Due to the long-range Coulomb interaction, the electrostatic potential 
change is determined by the transfered charge throughout the 
metal-SWNT-junction (Eq. \ref{VES}). For a given metal-SWNT 
distance $\Delta L$, its magnitude in the middle of the SWNT increases 
with the increasing SWNT size although the charge transfer is 
small except at the several layers immediately adjacent to the electrodes. 
The magnitude of the potential change in the interior of the SWNT 
saturates at the same length where the finite SWNT approaches the 
bulk limit, i.e., $50$ unit cells corresponding to a length of $21.1(nm)$, 
for both Au-NT-Au and Ti-NT-Ti junctions. For a given metal-SWNT 
distance $\Delta L$, the magnitude of the potential shift at the metal-SWNT 
interface is approximately constant for all the finite SWNTs studied. 

The contact-induced charge transfer processes are often characterized 
as ``charge-tranfer doping''. If we follow the common practice in 
literature, the SWNT is  ``hole-doped'' by contacting to gold (high 
work function) and ``electron-doped'' by contacting to titanium 
(low work function) electrode. 
Here it is important to recognize the difference in the physical processes 
governing the short-range and long-range electrostatics of the 
metal-SWNT interface. The charge transfer close to the metal-SWNT 
interface reflects the bonding configuration change upon contact to the 
metallic surfaces, which cannot contribute directly to transport since the 
corresponding charge distribution is localized~\cite{XueMol1,XueMol2}. 
Moving away from the interface, the effect due to the metal-SWNT 
coupling is reduced. For the longer SWNT molecule 
which has approached the bulk limit, the effect of the interface coupling 
on the electron states in the middle of the SWNT can be essentially 
neglected. However, since the electron occupation is determined by the 
Fermi-Dirac distribution of the metallic electrodes, the charging state 
in the interior of the SWNT which has approached the bulk limit 
is determined by the lineup of the SWNT bands relative 
to the metal Fermi-level, which in turn is determined by the self-consistent 
potential shift across the metal-SWNT-metal junction. Within the coherent 
transport regime, the transfered charge in the interior affects current 
indirectly by modulating the potential landscape acrosss 
the metal-SWNT-metal junction, which determines the electron 
transmission coefficient through Eq. \ref{TEV}.  

\begin{figure}
\includegraphics[height=3.2in,width=5.0in]{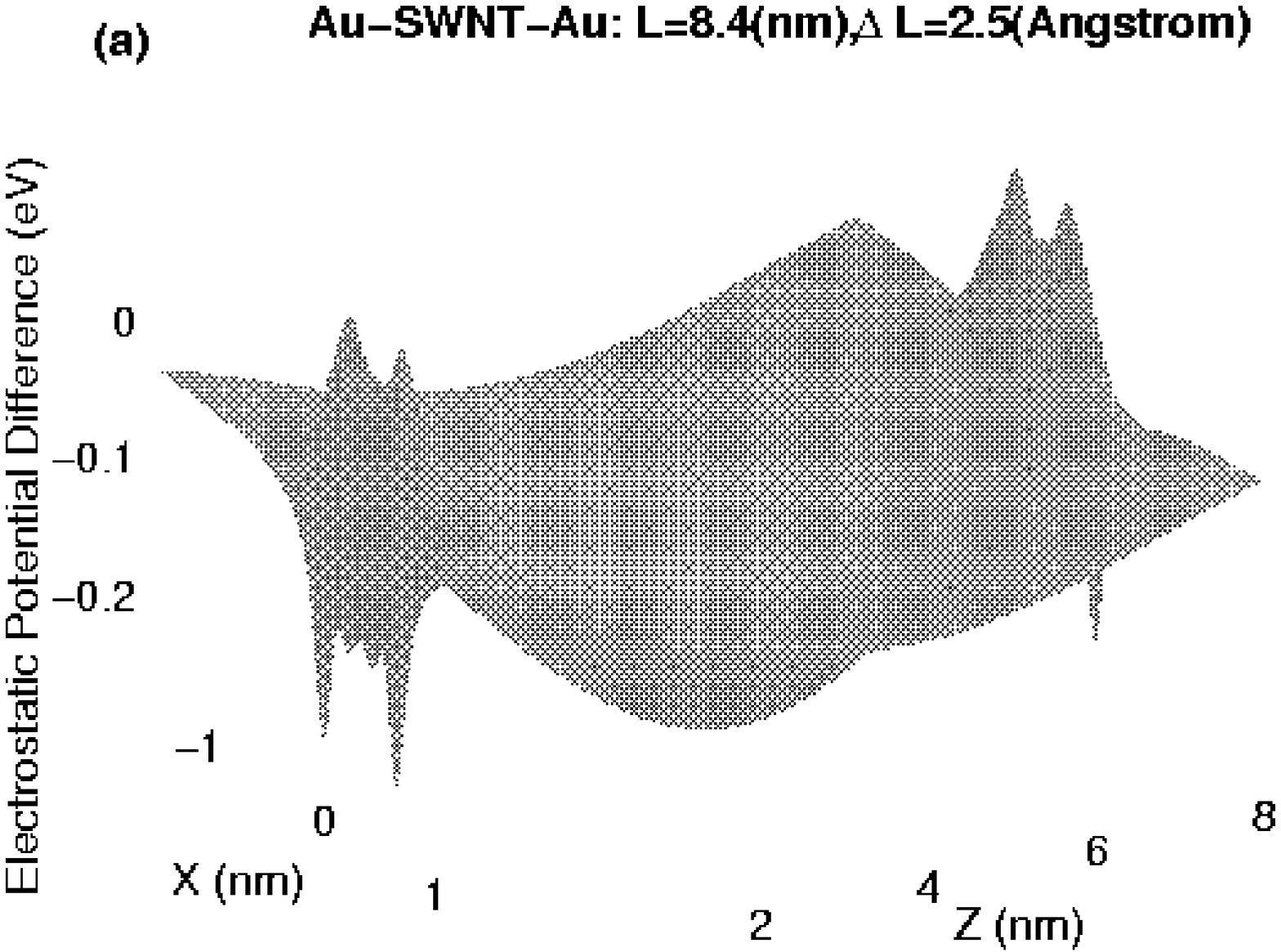}
\includegraphics[height=3.2in,width=5.0in]{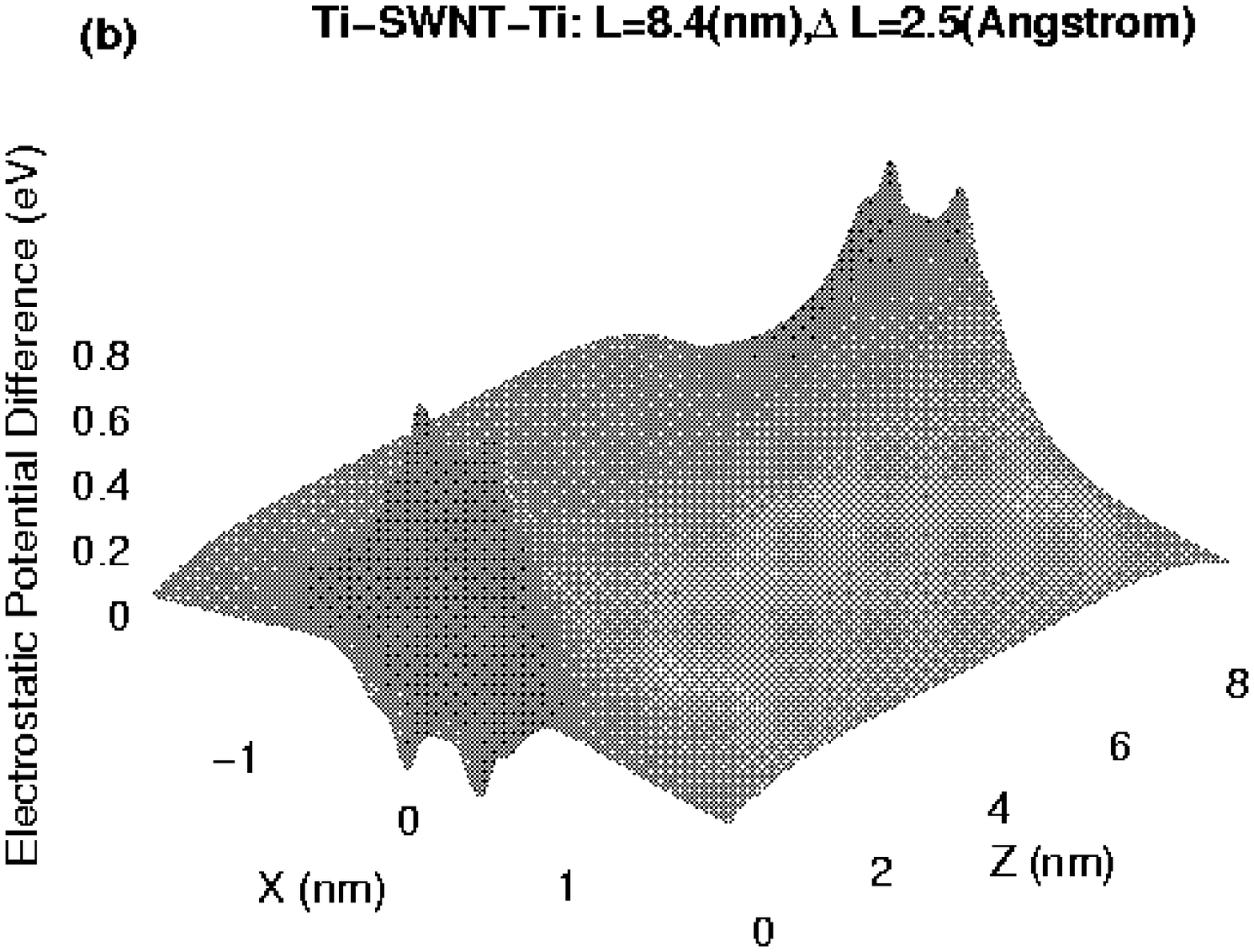}
\caption{\label{xueFig9} (Color online) Cross sectional view of electrostatic 
potential change at the Au-SWNT-Au (upper figure) and Ti-SWNT-Ti 
junction (lower figure) for SWNT molecule length 
of $8.4(nm)$ and metal-SWNT distance of $\Delta L=2.5(\AA)$. The 
SWNT diameter is $0.78(nm)$. The electrostatic potential change shown 
here is induced by the charge transfer across the interface and calculated using 
Eq. \ref{VES}. }
\end{figure}
   
A common feature of previous theoretical work on carbon nanotube 
devices is the use of the electrostatics of 
an ideal cylinder,~\cite{Xue99NT,TersoffNT,Odin00,De02,NTLimit} which 
neglects the electrostatic potential variation across the narrow region around 
the cylindrical surface where the $\pi$-electron density is 
non-negligible. However, the 
electrostatics of \emph{any nanostructure is three-dimensional}. For the 
cylindrical SWNT, this means that the electrostatic potential across the SWNT 
junction varies both parallel and perpendicular to the NT axis and on the 
atomic-scale. This is clearly seen from the three-dimensional plot of the 
electrostatic potential change in Fig.\ \ref{xueFig9}. 
For the $(10,0)$ SWNT with a diameter of $\approx 0.8(nm)$, the variation 
of the charge transfer-induced electrostatic potential change inside the 
SWNT cylinder is small, but decays to about $1/4$ of its value at the 
cylindrical center $1(nm)$ away from the SWNT surface for both 
the Au-SWNT-Au and Ti-SWNT-Ti junctions. 

\begin{figure}
\includegraphics[height=3.2in,width=5.0in]{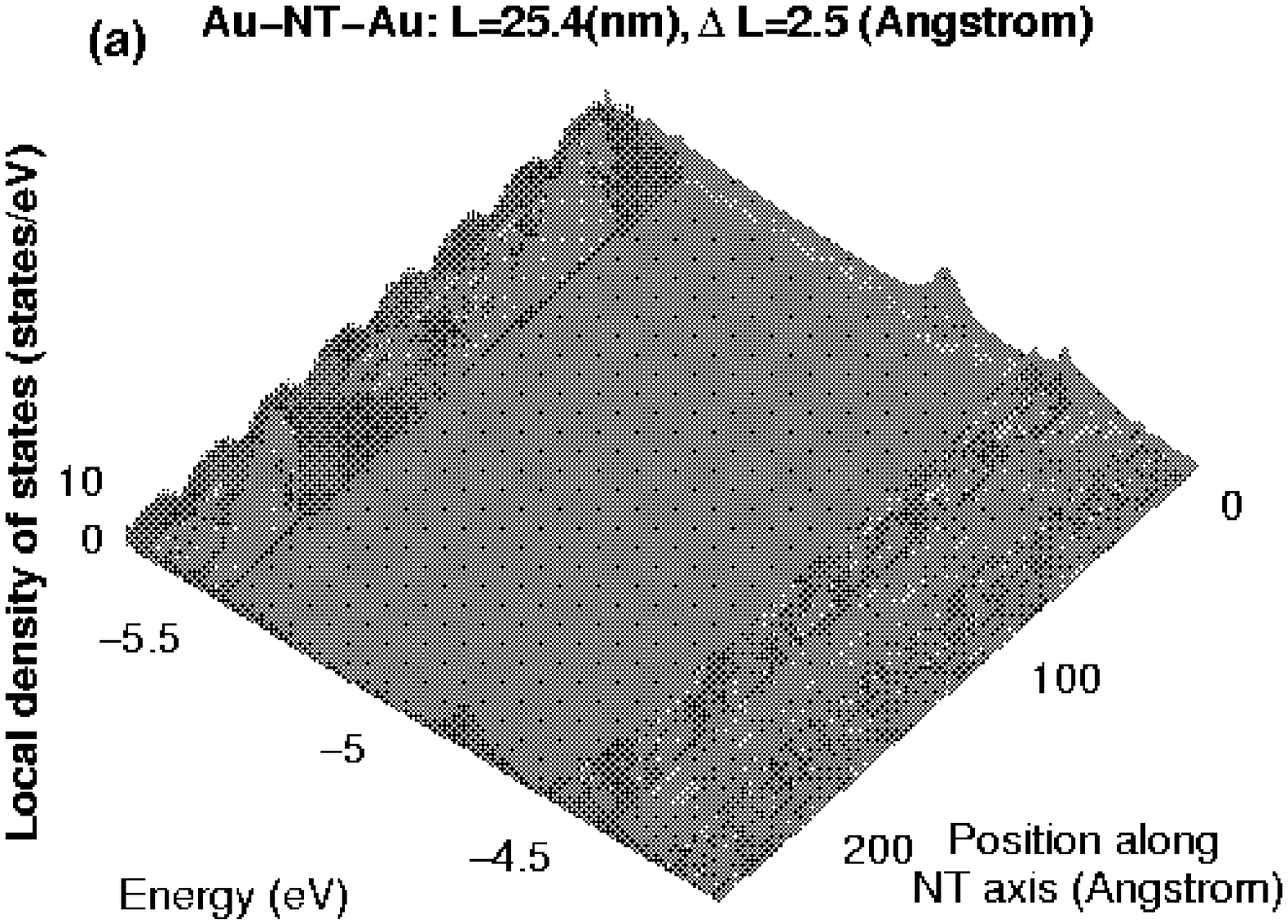}
\includegraphics[height=3.2in,width=5.0in]{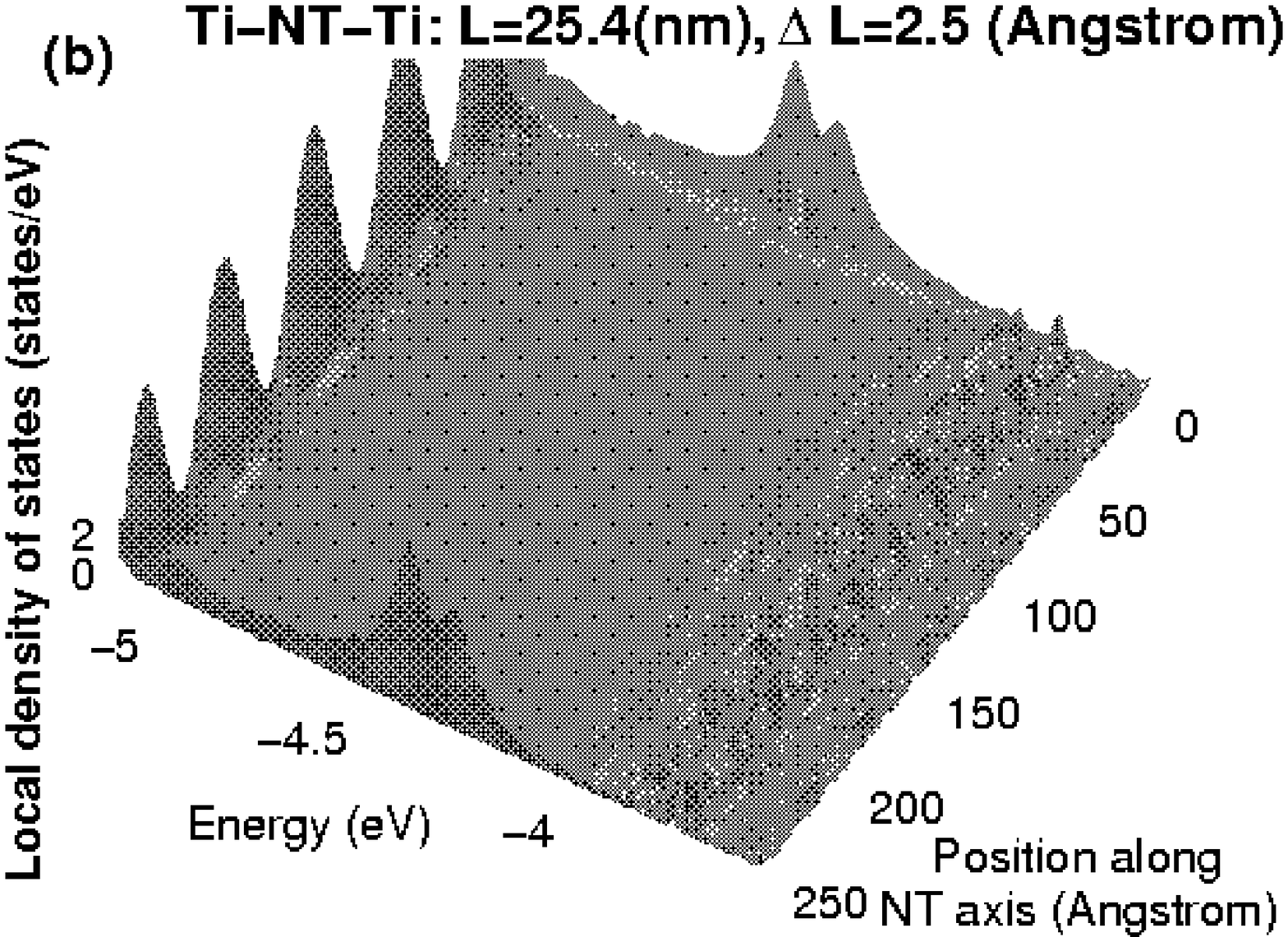}
\caption{\label{xueFig10} (Color online)
Three-dimensional plot of the local density of states at the Au-SWNT-Au (a) 
and Ti-SWNT-Ti (b) junctions as a function of position along the NT axis for 
SWNT length of $25.4(nm)$ and metal-SWNT distance of 
$\Delta L=2.5(\AA)$. Note that the sharp peaks around $ -4.5(eV)$ due to 
the dangling bond state at the ends of the isolated SWNT 
(Fig. \ref{xueFig4}(b)) have been replaced by broadened peaks within the 
band gap due to the MIGS at the metal-SWNT molecule interface. }
\end{figure}

The confined cylindrical geometry and three-dimensional electrostatics of the 
metal-SWNT interface lead to a profound change in the physical picture 
of the band shift, which applies to both finite SWNT molecules and long 
SWNT wires.~\cite{XueNT03,XueNT04} In particular, the shift of the local 
density of states along the nanotube axis \emph{does not follow} the 
change in the electrostatic potential along the nanotube axis, 
although this is commonly assumed 
in the literature. This is illustrated in the three-dimensional plot of 
the LDOS as a function of position along NT axis in Fig.\ \ref{xueFig10}. 
Note that although the electrostatic potential varies by an amount 
$\geq 0.5(eV)$ going from the metal-SWNT interface to the middle of 
the $60$-unit cell SWNT molecule for both junctions (Fig.\ \ref{xueFig7}), 
there is almost no shift of the conduction and valence band edge going from 
the interface to the middle of the SWNT molecule. This is in contrast 
with the planar metal-semiconductor interface, where the band shift 
away from the interface dipole layer follows the electrostatic potential 
change since it varies only in one direction \emph{and} 
on a length scale large compared to the corresponding unit cell size.  

The lack of connection between band shift and electrostatic potential 
change along the SWNT axis is obvious considering the 3-d nature of the 
electrostatics: Since the electrostatic potential change varies strongly 
in the direction perpendicular to the SWNT axis where the carbon 
pi-electron density is significant, there is no simple connection between 
the band shift and the electrostatic potential change at the cylindrical 
surface of the SWNT or at any other distance away from the SWNT axis. 
The relevant physics can be understood as follows: For the nanoscale 
SWNT considered here, the molecular-size interface dipole induces a 
long-range three-dimensional electrostatic potential change of  
$\sim 0.5 (eV)$ within $\sim 5(nm)$ of the interface, which is much 
weaker than the atomic-scale electrostatic potential variation within the 
bare SWNT. Since the LDOS of the SWNT junction is obtained from 
the Hamiltion corrected by the charge transfer-induced electrostatic 
potential change, we can expect the effect on the spatial variation 
of the LDOS away from the interface due to such correction is small 
compared to the strong atomic-scale potential variations included 
implicitly in the initial Hamiltonian $H_{0}$. The effect of the 
electrostatic potential change on the LDOS in regions within 
$\sim 5(nm)$ of the metal-SWNT interface is thus similar to that 
of small molecules in molecular tunnel junctions, where detailed 
studies in Ref. \onlinecite{XueMol2} have shown that the 
charge transfer-induced electrostatic potential change 
in the molecular junction doesn't lead to a rigid shift of the molecular 
energy levels (or band edges), but can have different effects on different 
molecular states (or band structure modification) depending on their 
charge distributions.
   
\subsection{``Band'' lineup and electron transmission across the 
metal-finite SWNT molecule interface}

\begin{figure}
\includegraphics[height=3.2in,width=4.0in]{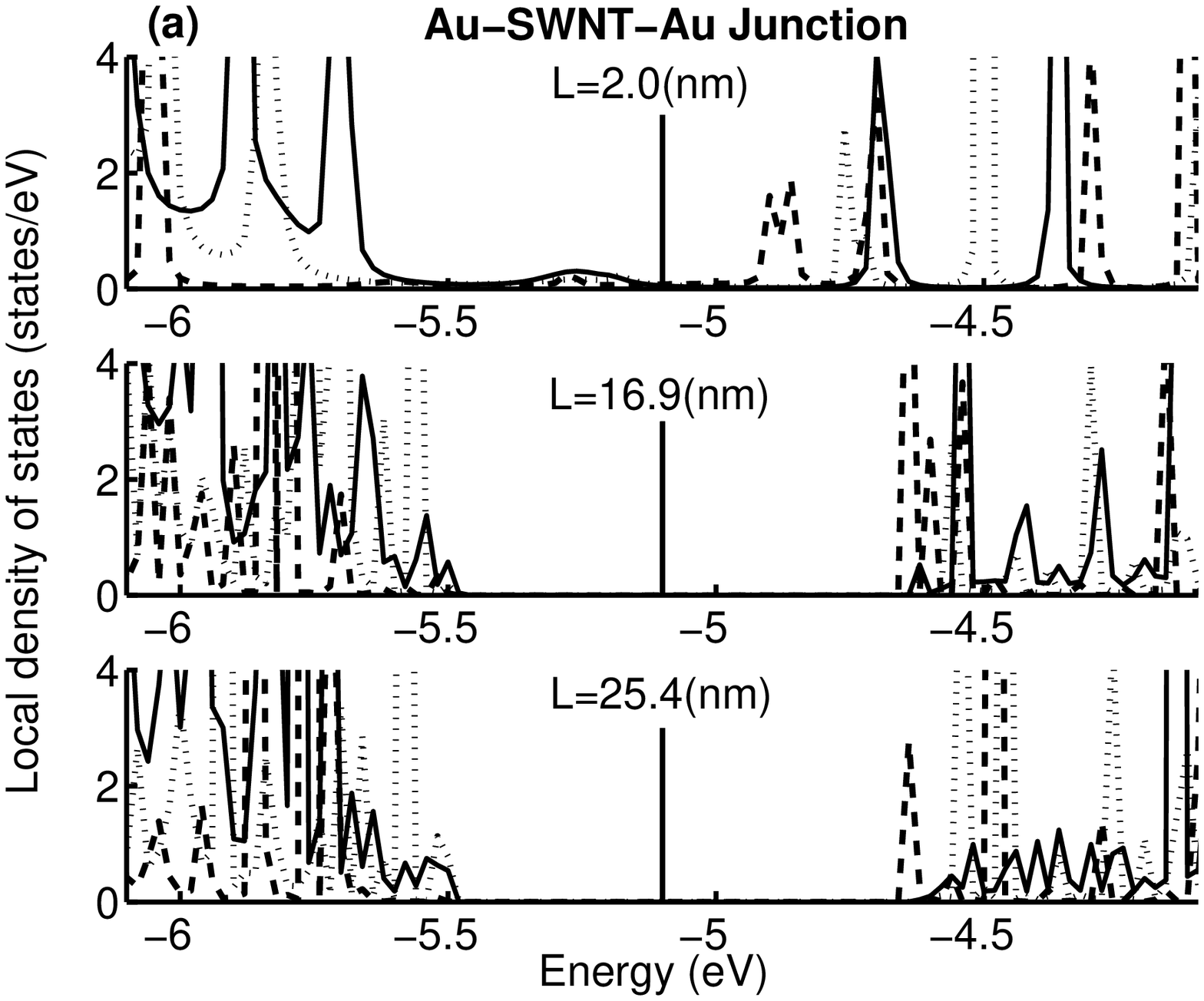}
\includegraphics[height=3.2in,width=4.0in]{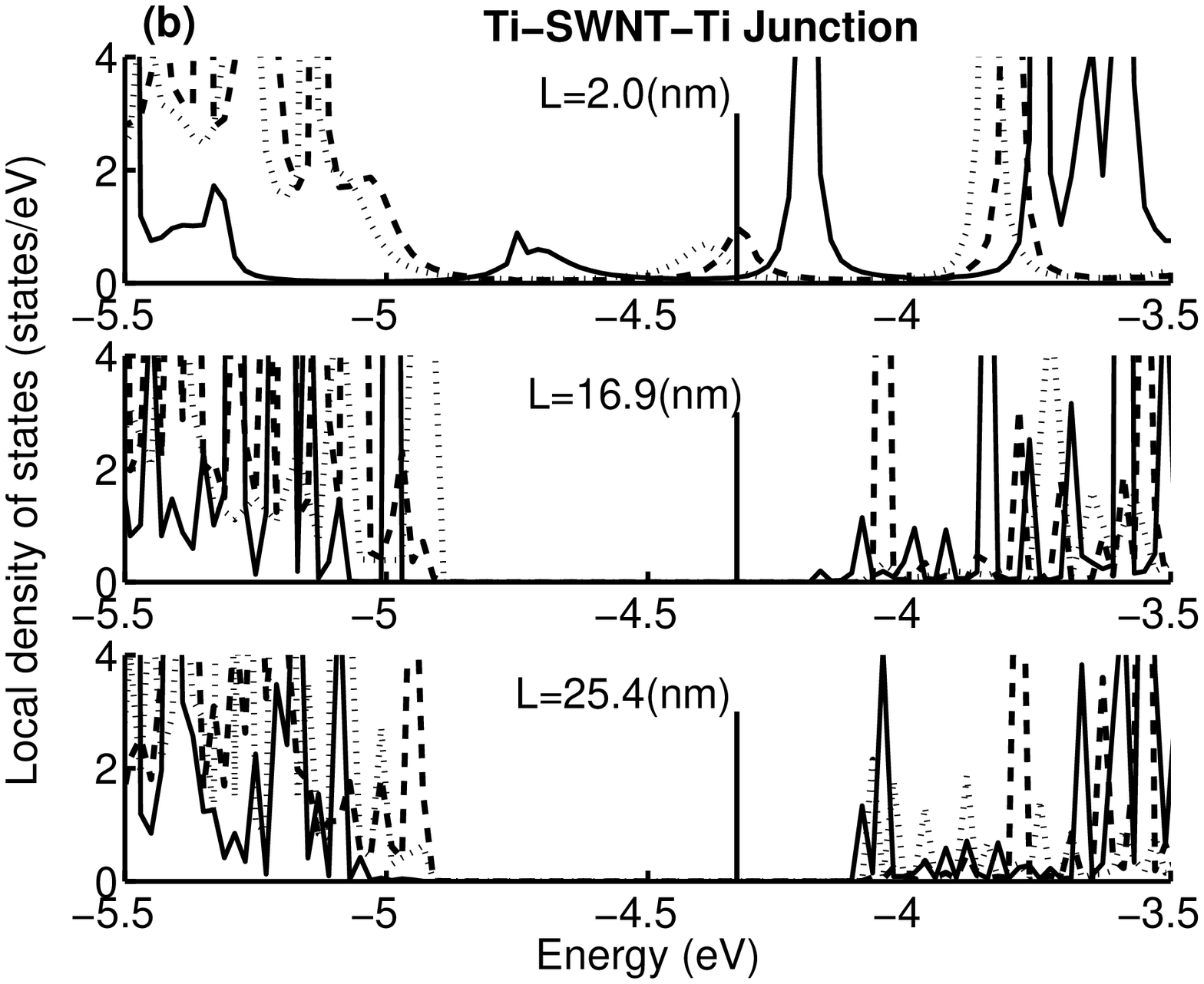}
\caption{\label{xueFig11}
Local density of states at the middle of the Au-SWNT-Au junction (a) and 
Ti-SWNT-Ti junction (b) for SWNT length of $2.0,16.9$ and 
$25.4(nm)$ respectively. Solid line: $\Delta L=2.0(\AA)$. Dotted line: 
$\Delta L=2.5(\AA)$. Dashed line: $\Delta L=3.0(\AA)$. The vertical lines 
show the position of the metal Fermi-level. }    
\end{figure}

For a planar metal-semiconductor interface, the band lineup is determined 
once the electrostatic potential drop across the interface is known. 
The horizontal lines in the potential plots of Figs. 
\ref{xueFig6}-\ref{xueFig8}(b) denote the work function differences 
between the electrodes and the bulk SWNT. For a bulk metal-semiconductor 
interface, this would have given the magnitude of the potential shift 
which aligns the Fermi-level across the interface. But for the metal-finite 
SWNT interface considered here, the band lineup should be determined 
from the local density of states (LDOS) in the middle of the SWNT. This 
is shown in Fig. \ref{xueFig11} for both Au-SWNT-Au and Ti-SWNT-Ti 
junctions.  

\begin{figure}
\includegraphics[height=3.2in,width=4.0in]{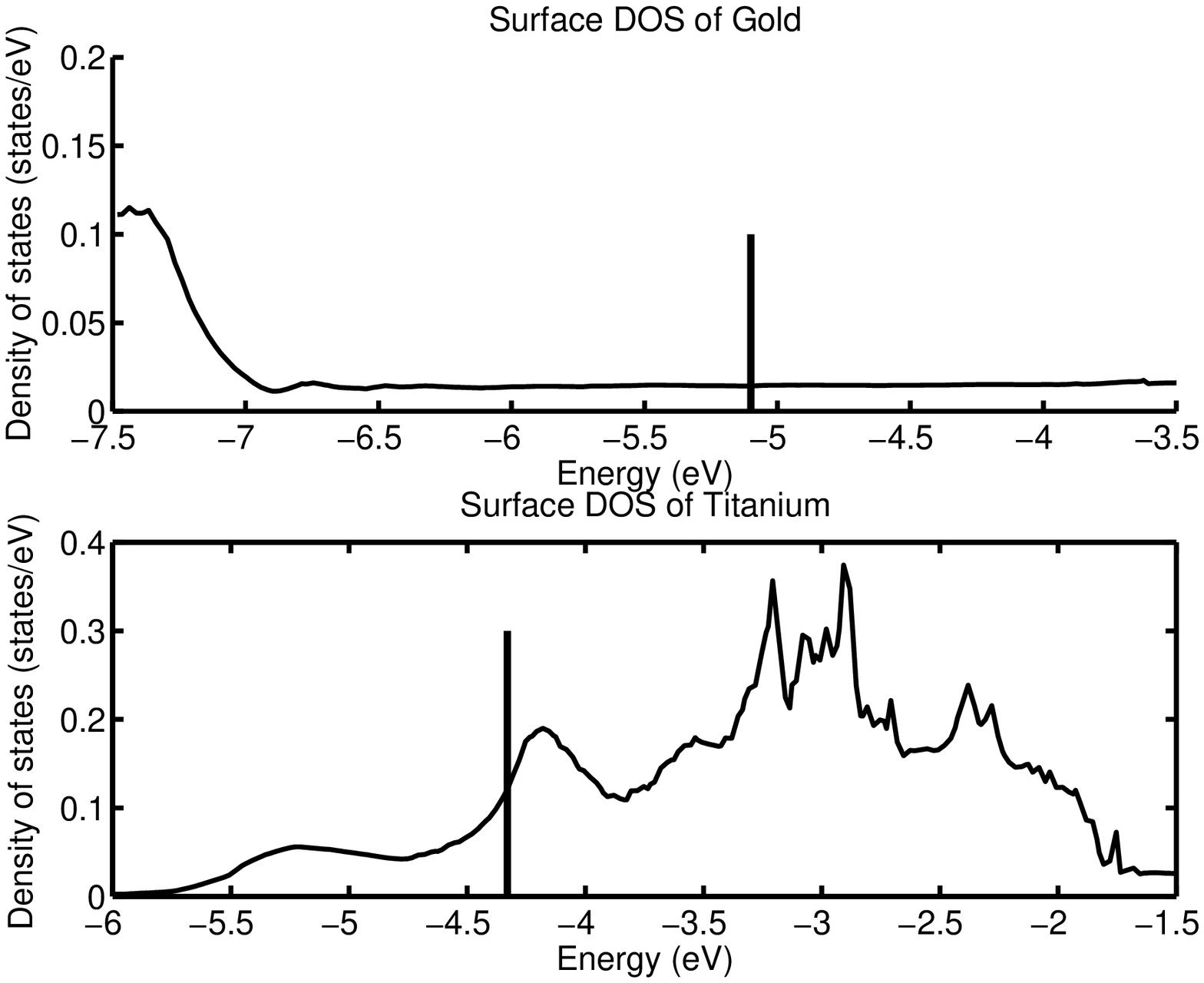}
\caption{\label{xueFig12}
Surface density of states of the gold and titanium electrodes. The vertical 
lines show the position of metal Fermi-level. }    
\end{figure}

\begin{figure}
\includegraphics[height=3.2in,width=4.0in]{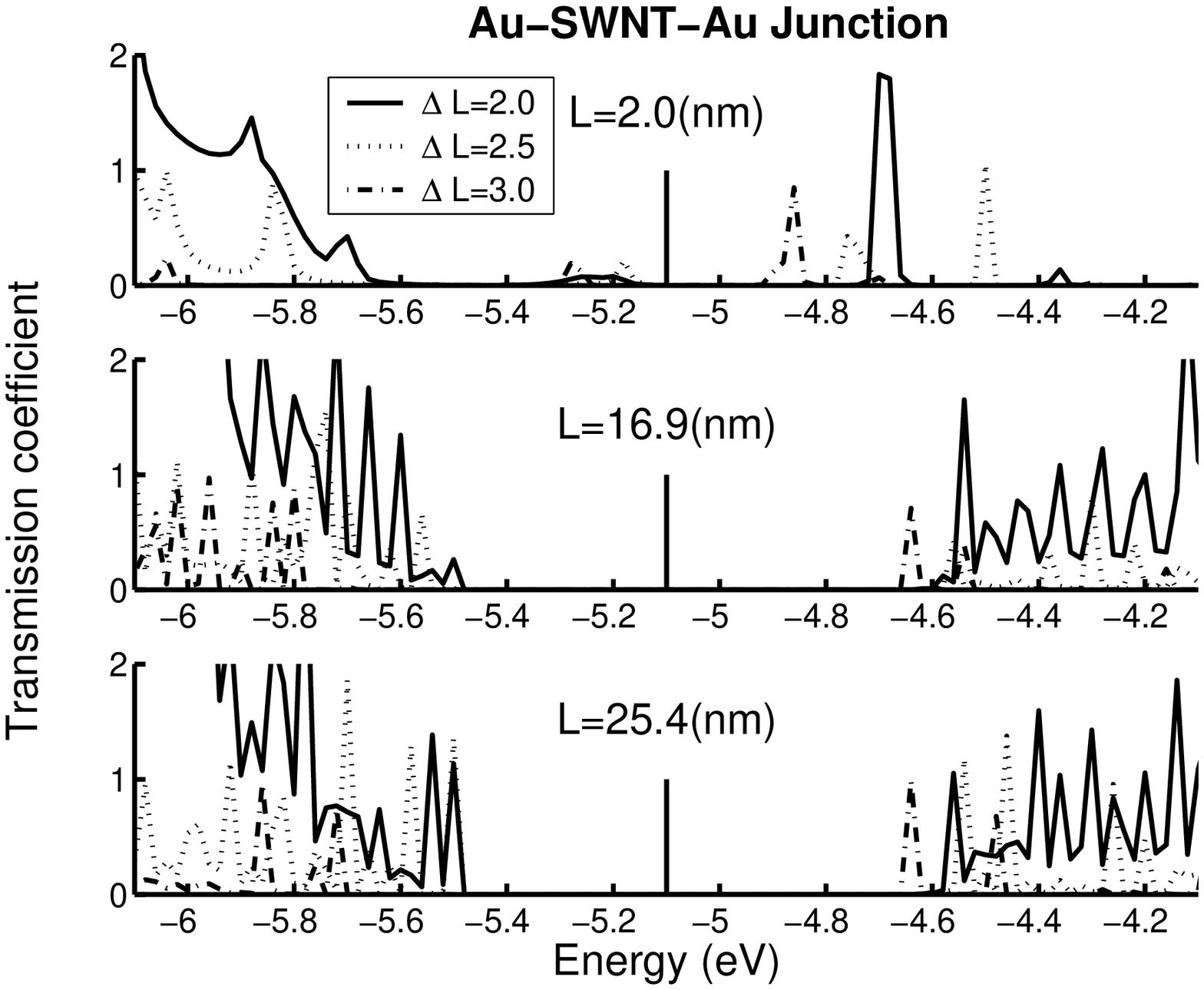}
\includegraphics[height=3.2in,width=4.0in]{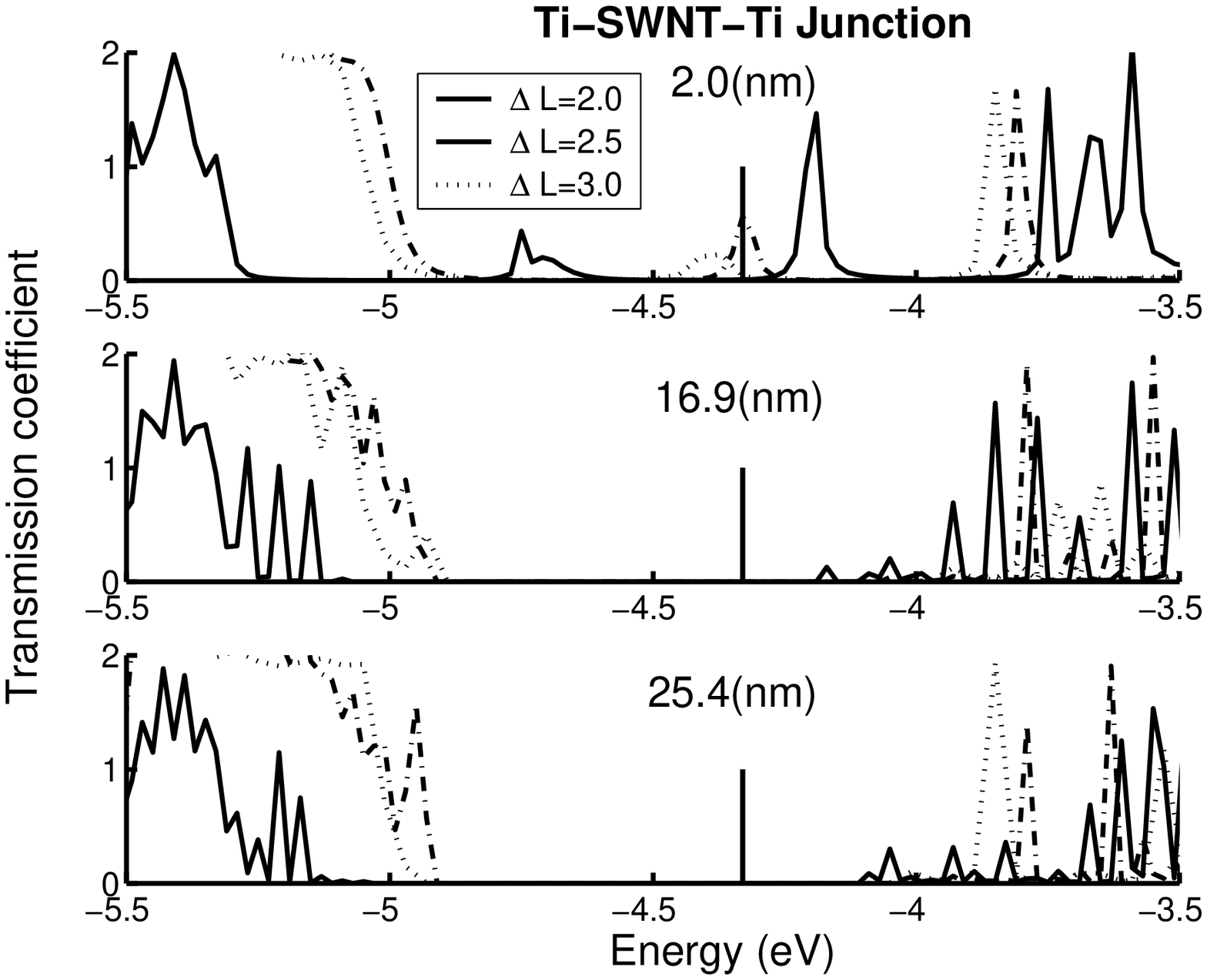}
\caption{\label{xueFig13}
Electron transmission characteristics of the Au-SWNT-Au (upper figure) 
junction and Ti-SWNT-Ti (lower figure) junction for SWNT length 
of $2.0,16.9$ and $25.4(nm)$ and metal-SWNT distance of $2.0,2.5$ and 
$3.0 (\AA )$ respectively. The vertical lines show the 
position of the metal Fermi-level at each junction.  }
\end{figure}

The ``band'' lineup relevant to the transport characteristics can also 
be determined equivalently from the electron transmission characteristics 
of the equilibrium metal-SWNT-metal junction, which is calculated using 
Eq. \ref{TEV} and depends on the surface electronic structure, the 
coupling across the interface and the electronic structure of the SWNT 
molecule. The surface density of states of the bare gold and 
titanium electrodes calculated using tight-binding parameters~\cite{Papa86} 
are shown in Fig. \ref{xueFig12}, while the  
transmission characteristics of the metal-SWNT-metal junctions are 
shown in Fig. \ref{xueFig13}. For the shortest SWNT in the molecular limit 
($2.0 nm$), there is significant transmission around the metal Fermi-level 
$E_{f}$ which is suppressed rapidly with increasing SWNT length. The 
difference in the electron transmission through the SWNT conduction band 
region in the Au-SWNT-Au and Ti-SWNT-Ti junctions is mostly due to the 
difference in the electrode band structures above $E_{f}$ ($sp$-band for Au 
and $d$-band for Ti).      
 
From both the LDOS and transmission 
characteristics of the $60$-unit cell SWNT, we can determine that for 
the Au-SWNT-Au junction the Fermi-level location goes from slightly 
below (by $\sim 0.1$ eV) the mid-gap of the $60$ unitcell SWNT to 
the mid-gap as the gold-SWNT distance increases from $2.0(\AA)$ 
to $3.0(\AA)$. For the Ti-SWNT-Ti junction, the Fermi-level location 
goes from above (by $\sim 0.25 $ eV) the mid-gap of the $60$-unit cell 
SWNT molecule to the midgap as the titanium-SWNT distance increases 
from $2.0(\AA)$ to $3.0(\AA)$. Note that this value is approximately the 
same for SWNTs longer than $40$-unit cell ($16.9$ nm), i.e., the same 
length where the magnitude of the electrostatic potential change in the 
middle of the SWNT begins to saturate 
(Figs. \ref{xueFig6}-\ref{xueFig8}(b)).   

The physical principles of Schottky barrier formation at the 
metal-SWNT molecule interface can thus be summarized as follows: 
Since the effect of the interface perturbation on the electron states 
inside the SWNT molecule is small, for the SWNTs that are long enough 
to approach the bulk limit, the metal Fermi-level position should be 
close to the middle of the gap since otherwise extensive charge transfer 
will occur inside the SWNT junction. Since the screening of the work function 
difference inside the SWNT junction is weak, the metal Fermi-level should 
be below (above) the middle of the gap for a high (low) workfunction 
metal so that the net decrease (increase) of electrons inside the SWNT 
molecule shifts the SWNT band edge down (up) relative to the metal 
Fermi-level. Exactly how this is achieved from the interface to the middle 
of the channel will depend on the details of the contact (types of metal and 
strength of interface coupling). At the weak coupling limit, the lineup of 
the Fermi-level for the SWNT molecules which have reached the bulk 
limit is such that the perturbation of the electron states inside the SWNT 
molecule is minimal, i.e., at mid-gap.  Note that since the LDOS around 
the midgap is negligible inside the SWNT, the magnitude of the transfered 
charge in the middle of the SWNT molecule is approximately independent 
of the interface coupling strength despite the different band lineup scheme at 
three different metal-SWNT distances (Figs. \ref{xueFig6}-\ref{xueFig8}(a)).      

\section{Length and temperature dependence of the conductance of the 
metal-SWNT molecule interface}

\begin{figure}
\includegraphics[height=3.2in,width=3.6in]{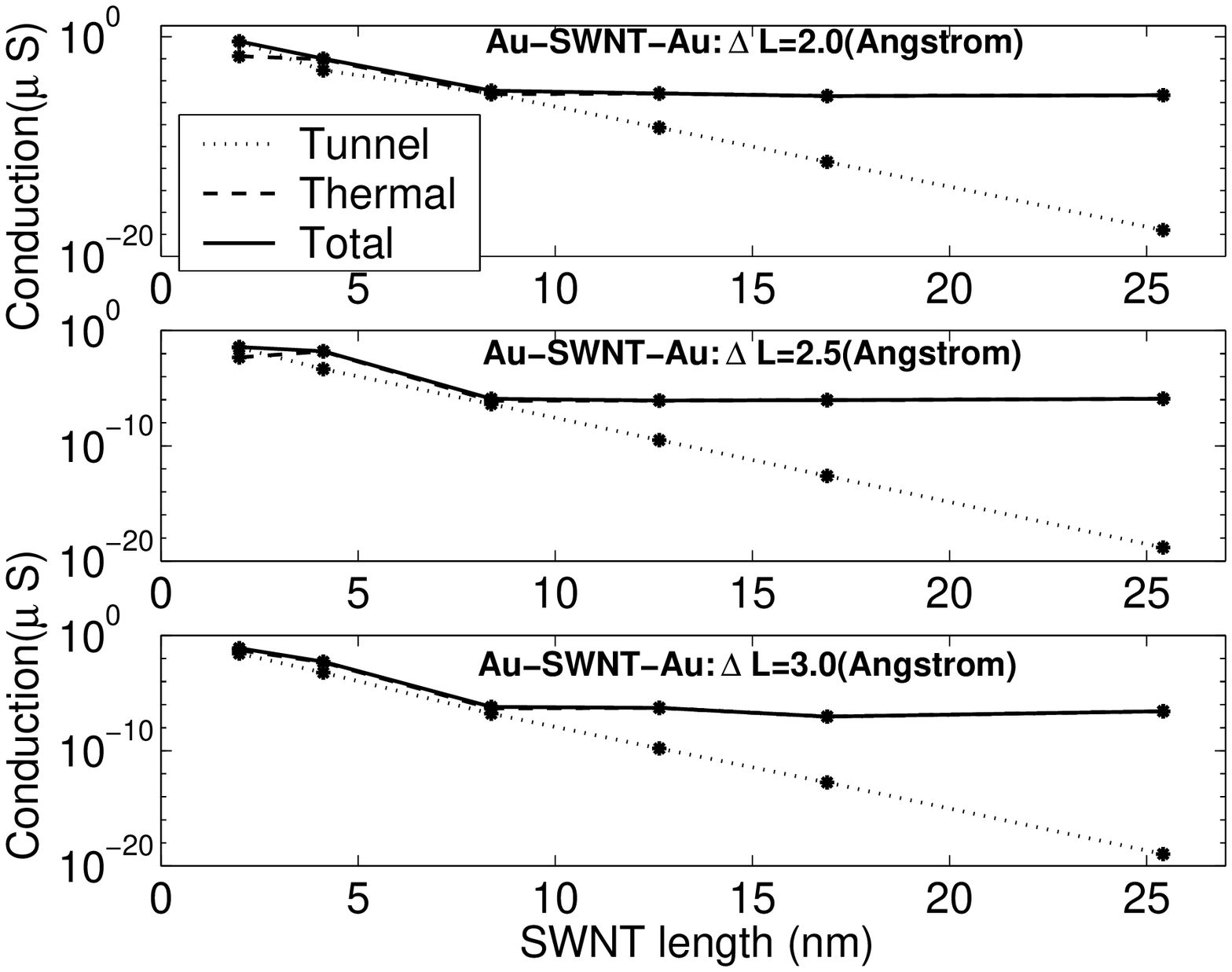}
\includegraphics[height=3.2in,width=3.6in]{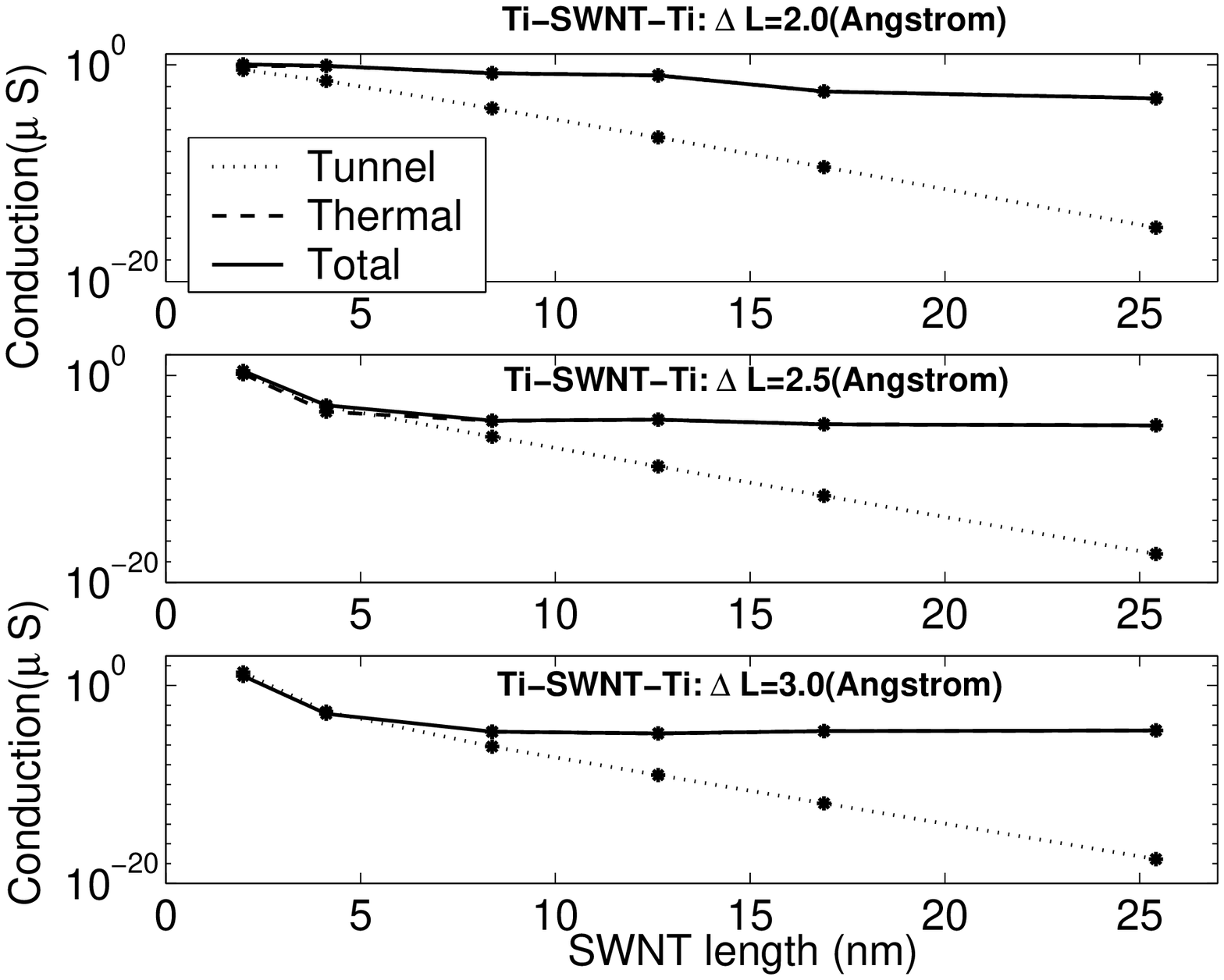}
\caption{\label{xueFig14}
Room temperature conductance of the Au-SWNT-Au (upper figure) 
junction and Ti-SWNT-Ti (lower figure) junction as a function of the 
SWNT length at three different metal-SWNT distances. }
\end{figure}

Given the electrostatic potential change $\Delta V$ across the 
metal-SWNT interface, we can caculate the length and temperature 
dependence of the metal-SWNT-metal junction conductance using 
Eq. \ref{GT}. 
The length dependence of the junction conductance at room temperature 
is shown in Fig. \ref{xueFig14} for both Au-SWNT-Au and Ti-SWNT-Ti 
junctions at the three metal-SWNT distances. We have separated the junction 
conductance into the tunneling and thermal-activation contributions as 
discussed in sec. \ref{Theory}.  

\begin{figure}
\includegraphics[height=3.2in,width=4.0in]{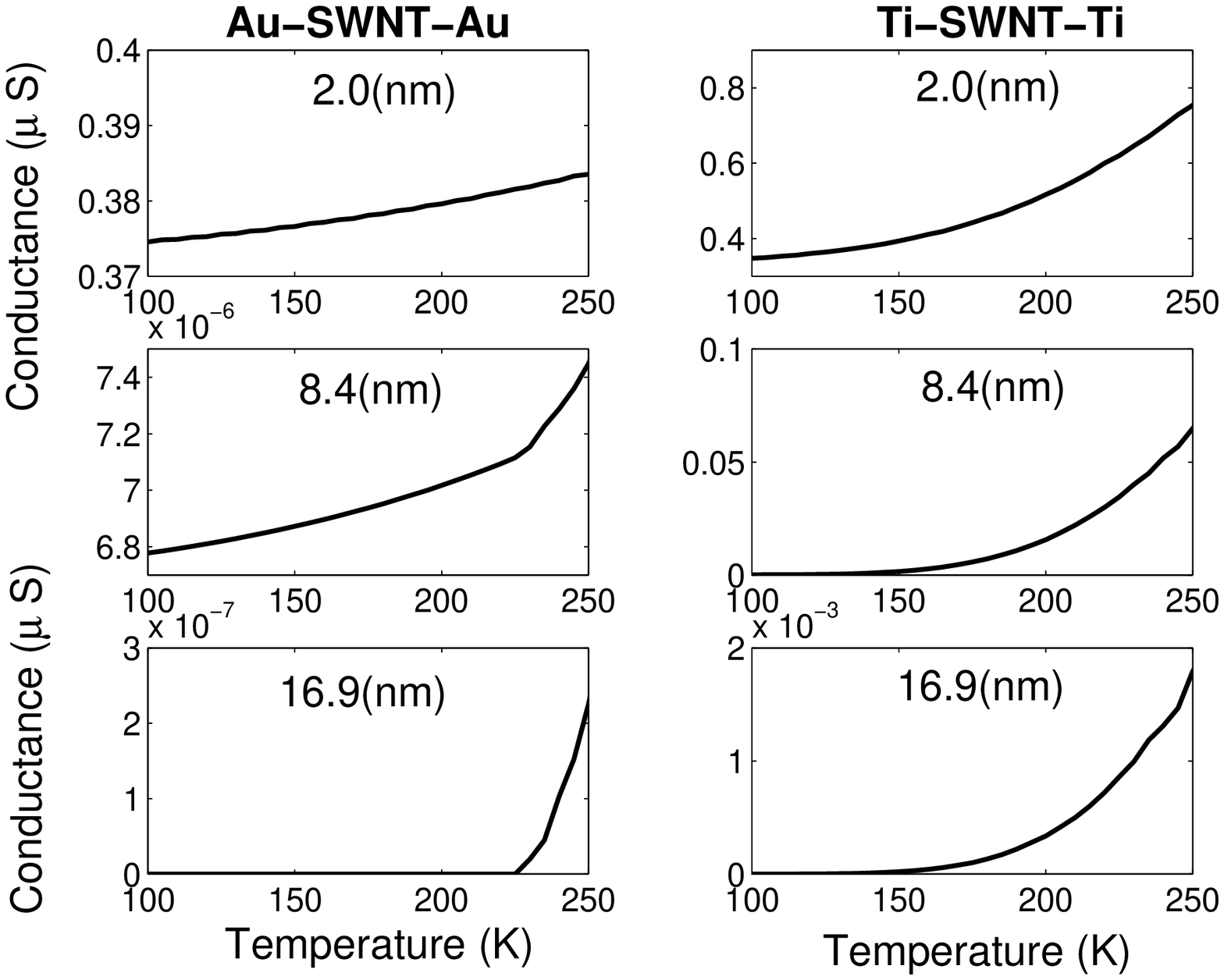}
\caption{\label{xueFig15}
Temperature dependence of the conductance of the Au-SWNT-Au 
(left figure) junction and Ti-SWNT-Ti (right figure) junction as a function 
of the SWNT length at metal-SWNT distance of $\Delta L=2.0 (\AA)$. }
\end{figure}

The tunneling conductance  (also the zero-temperature conductance) 
for both junctions decreases exponentially with the SWNT length for 
SWNT longer than $4.1(nm)$ (Fig. \ref{xueFig14}), where the 
perturbation of the electron states inside the SWNT due to the interface 
coupling can be neglected. The exponential decay with length for 
tunneling across a finite molecular wire in contact 
with two metal electrodes has been analyzed in detail in recent literature 
using either simple tight-binding theory~\cite{Joachim} or complex 
band structures calculated from first-principles theory.~\cite{Sankey} 
But the essential physics can be captured from the simple WKB picture of 
tunneling through potential barriers with constant barrier height. 
A separation of the contact and molecule core effect 
on the tunneling resistance can thus be achieved using the functional 
relation $R=R_{0}e^{dL}$, where $R_{0}$ is the contact resistance 
and $d$ is the inverse decay length for tunneling across the SWNT molecule. 
We find that the Au-SWNT-Au 
junction has the contact resistance $R_{0}=0.115, 1.88, 2.59 (M\Omega)$ 
and inverse decay length of $d=1.68,1.68,1.68(1/nm)$ for the Au-SWNT 
distance of $\Delta L=2.0,2.5,3.0 (\AA)$ respectively. The Ti-SWNT-Ti 
junction has  the contact resistance $R_{0}=0.023, 3.14, 4.95 (M\Omega)$ 
and inverse decay length of $d=1.51,1.52,1.53(1/nm)$ for the Ti-SWNT 
distance of $\Delta L=2.0,2.5,3.0 (\AA)$ respectively. Note that the 
contact resistance increases rapidly with the increasing metal-SWNT distance 
due to the reduced interface coupling, but the inverse decay length (which 
is a bulk-related parameter) remains approximately constant.~\cite{Sankey} 
The total conductance of the metal-SWNT-metal junction at room temperature 
saturates with increasing SWNT length. This is due to the fact that the 
potential shift extends over a range comparable to the half of the SWNT 
length until the SWNT reaches the bulk limit. For longer SWNT, the 
tunneling is exponentially suppressed while the transport becomes 
dominated by thermal activation over the potential barrier whose 
height is approximately constant for all the SWNTs investigated. 

The length and temperature dependence of the metal-finite SWNT-metal 
junction can also be seen more clearly from Fig. \ref{xueFig15}, where 
we show the conductance of the SWNT junction as a function of temperature 
for lengths of $2.0$, $8.4$ and $16.9$ (nm) in both Au-SWNT-Au and 
Ti-SWNT-Ti 
junctions and in the strong coupling limit ($\Delta L=2.0\AA$). For the 
shortest SWNT molecule ($2.0$ nm) studied, both tunneling and 
thermal contributions to the conductance at room-temperature 
are significant. So the condutance increases only by a factor of 2 going 
from $100(K)$ to $250(K)$ for the Ti-SWNT-Ti junction 
and is almost temperature independent for the Au-SWNT-Au junction. The 
thermionic-emission contribution begins to dominate over the tunneling 
contribution at SWNT length of $8.4$ (nm) and longer, correspondingly the 
increase of conductance with temperature is faster. But overall the 
temperature dependence is much weaker than the exponential dependence 
in, e.g., electron transport through the planar metal-semiconductor 
interfaces.~\cite{MS}   

The length and temperature dependence of the SWNT molecule junction can 
be understood rather straightforwardly using the Breit-Wigner 
formula,~\cite{Landau} first introduced by Buttiker~\cite{ButtikerRTD} for 
electron transmission through double-barrier tunneling structures. 
For electron transmission within the energy gap between the 
highest-occupied-molecular-orbital (HOMO) and 
lowest-unoccupied-molecular-orbital (LUMO) of the SWNT molecule, we 
can approximate the energy dependence of the transmission coefficient as  
\begin{equation}
\label{BW}
T(E) \approx \sum_{i=HOMO,LUMO} \frac{\Gamma_{i;L}\Gamma_{i;R}}
       {(E-E_{i})^{2}+1/4(\Gamma_{i;L}+\Gamma_{i;R})^{2}} 
\end{equation}   
where $\Gamma_{i;L(R)}$ (i=HOMO,LUMO) is the partial width of resonant 
transmission through the HOMO (LUMO) level due to elastic tunneling into 
the left (right) electrode respectively. Note that as the SWNT molecule 
reaches the bulk limit, the HOMO and LUMO levels give the valence band 
and conduction band edge respectively.  For given SWNT molecule and metallic 
electrodes, $\Gamma_{HOMO(LUMO);L(R)}$ is constant. The increase 
of transmission coefficient with energy from the Fermi-level $E_{f}$ towards 
the relevant band edge is thus of Lorentzian form, which is also generally 
true for nanostructures with only a finite number of conduction channels. 
From Eq. \ref{GT}, the temperature dependence of the conductance is thus 
determined by the tail of the Lorentzian around $E_{f}$ averaged over a 
range $\sim kT$  due to the thermal broading with the corresponding weight 
$\frac{df}{dE}(E-E_{f})=exp((E-E_{f})/kT)/(kT(exp((E-E_{f})/kT)+1)^{2})$. 
This leads to much weaker-than-exponential dependence on temperature of 
the junction conductance, as compared to the metal-semiconductor interface, 
where the exponential dependence of conductance on temperature is due to 
the exponential decrease of carrier densities with energy large enough to 
overcome the interface barrier.~\cite{MSBook} As the length of the SWNT 
molecule increases, the partial width $\Gamma_{HOMO(LUMO);L(R)}$ 
due to tunneling into the electrodes decreases exponentially 
(from the WKB approximation,~\cite{ButtikerRTD}) leading to the 
exponential dependence on junction length of the tunneling conductance.     

\section{Current-voltage characteristics of the 
metal-finite SWNT interface}

\begin{figure}
\includegraphics[height=3.2in,width=4.0in]{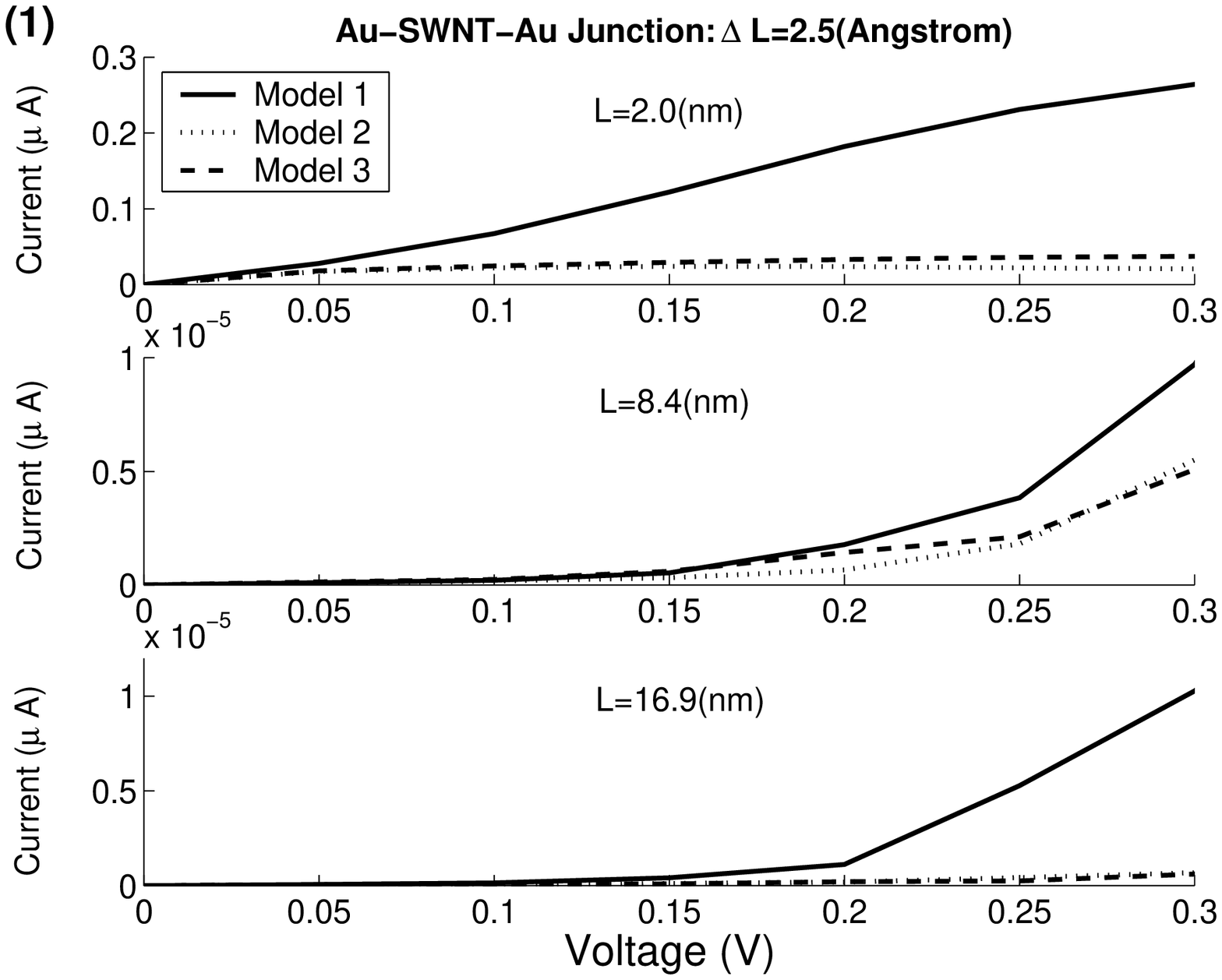}
\includegraphics[height=3.2in,width=4.0in]{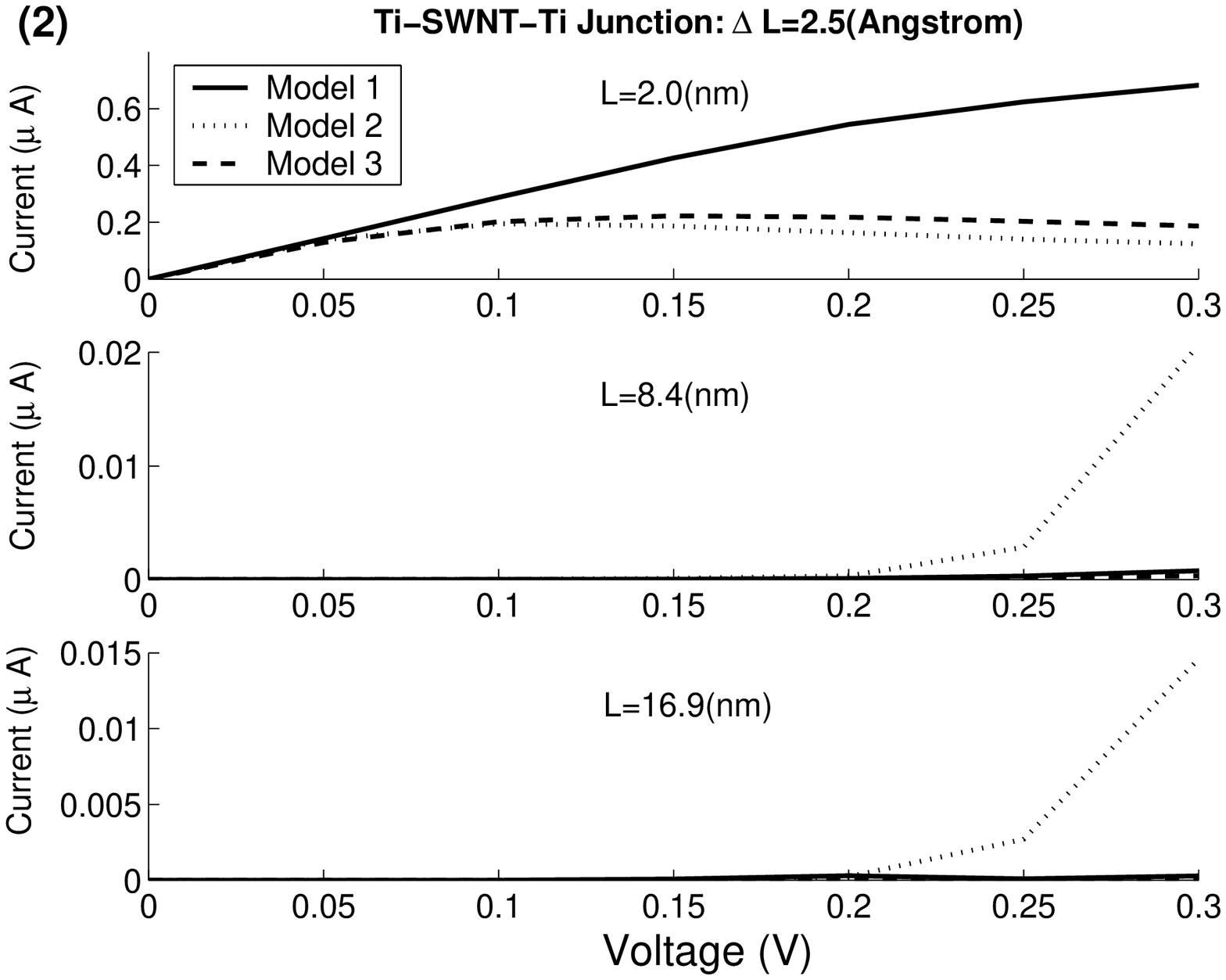}
\caption{\label{xueFig16}
Current-voltage characteristics of the Au-SWNT-Au (1) and the Ti-SWNT-Ti 
(2) junction for SWNT lengths of $2.0,8.4,16.9(nm)$ and metal-SWNT 
distance of $2.5(\AA)$. We consider three different models of 
electrostatic potential profile within the SWNT junction.  }
\end{figure}
  
In principle, to calculate the current-voltage characteristics of the 
metal-SWNT-metal junction, a self-consistent calculation of the charge and 
potential response will be needed at each bias voltage to take into account 
the screening of the applied electric field within the 
junctions.~\cite{XueMol2,Buttiker93} This is computationally 
demanding even for the self-consistent tight-binding method due to 
the large size of the SWNT molecule. Therefore, in this section we 
calculate the current-voltage characteristics using three different models 
of the electrostatic potential profiles in the metal-finite SWNT-metal 
junction in order to illustrate qualitatively the importance of the proper 
modeling of the self-consistent screening of the applied source/drain 
bias voltage.~\cite{XueMol2,Xue99Mol,DattaMol,RatnerMol} The fully 
self-consistent current transport is under investigation and will be 
reported in future publications. 

The three potential response models we choose are: (1) We assume all 
the voltage drop occurs at the metal-SWNT interface with the two 
interface contributing equally (Model 1); (2) We assume the voltage drop 
across the metal-SWNT-metal junction is piece-wise linear (Model 2); 
(3) We assume the voltage drops linearly across the entire metal-SWNT-metal 
junction (Model 3). The three potential models chosen here represent the 
source/drain field configuration at three different limits: In the absence 
of the SWNT molecule, we are left with the bare (planar) source/drain 
tunnel junction. For ideal infinitely conducting electrodes, the voltage 
drop will be linear with constant electric field across the source/drain 
junction. In general, sandwiching the SWNT molecule between the two 
electrodes leads to screening effect. If we neglect entirely the screening 
of the applied source/drain field by the SWNT molecule, we arrive at 
potential model 3. If the nanotube is infinitely conducting, we arrive at 
potential model 1. In practice, both the electrodes and the SWNT are 
not infinitely conducting, and the voltage drop can occur both across the 
metal-SWNT interface and inside the SWNT. Since the potential 
variation will be the largest close to the interface for the homogeneous 
SWNT assumed here, for model 2 we assume the potential profile is such 
that the magnitude of the field across the first unitcell of the SWNT at 
the two ends is $10$ times of that in the interior of the SWNT molecule. 
Note that we have neglected the electrostatic potential variation in the 
direction perpendicular to the source/drain field. For SWNTs with 
cylindrical structure, this can be important in a fully self-consistent 
analysis of the nonlinear current-voltage characteristics as we have 
seen in the previous sections. The three potential models chosen here 
are merely used to demonstrate the importance of the fully 
self-consistent study.     

\begin{figure}
\includegraphics[height=3.2in,width=5.0in]{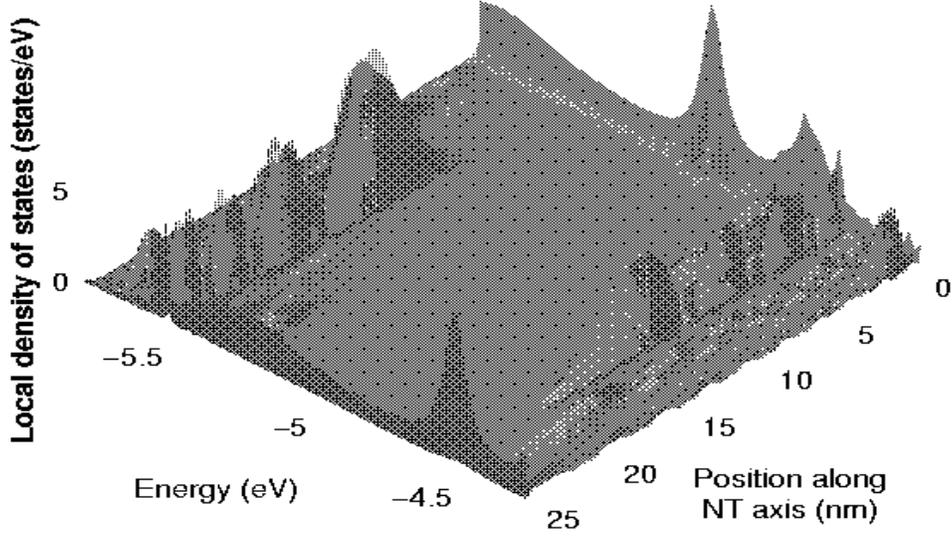}
\includegraphics[height=3.2in,width=5.0in]{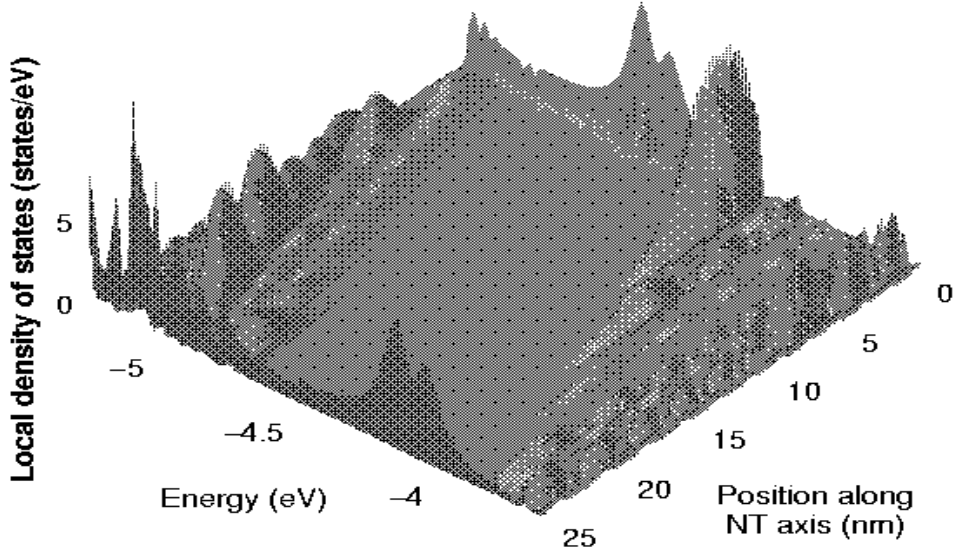}
\caption{\label{xueFig17} (Color online) Three-dimensional plot 
of the local density of states at the Au-SWNT-Au (a) and Ti-SWNT-Ti (b) 
junctions as a function of position along the NT axis for SWNT length 
of $25.4(nm)$ and metal-SWNT distance of $\Delta L=2.5(\AA)$ at 
source/drain bias voltage of $0.5 (V)$. We assume the voltage drops 
linearly across the SWNT junction (potential model 3). }
\end{figure}

The calculated current-voltage (I-V) characteristics of the 
metal-SWNT-metal junctions for SWNT lengths of $2.1,8.4,16.9(nm)$ and 
metal-SWNT distance of $2.5(\AA)$ are plotted in Fig. \ref{xueFig16} 
for both junctions. For electrostatic potential models 2 and 3, the I-V 
characteristics are obtained by superposing the assumed electrostatic 
potential profile onto the Hamiltonian of the equilibrium junction and 
evaluating its matrix element by direct numerical integration. We find 
that as the length of SWNT increases, the three different models of 
electrostatic potential response lead to qualitatively different 
current-voltage characteristics in both the magnitude of the current 
and its voltage dependence. This is because current transport is 
dominated by thermal-activation contribution for all the SWNT molecules 
investigated except the shortest ones. For the Au-SWNT-Au junction, 
we find that potential models 2 and 3 give qualitatively similar I-V 
characteristics, indicating that potential drop within the SWNT bulk 
is important. But for the Ti-SWNT-Ti junction, we find that potential 
models 1 and 3 give qualitatively similar I-V characteristics for SWNTs 
longer than $2.0(nm)$, indicating instead that potential drop across 
the metal-SWNT interface is important. 

The contact dependence of the source/drain field 
effect can also be seen more clearly by analyzing its effect on the SWNT 
electronic structure from Fig. \ref{xueFig17}, where we show the 
three-dimensional plot of the LDOS of the SWNT within the Au-SWNT-Au 
and Ti-SWNT-Ti junctions at applied bias voltage of $0.5 (V)$ and 
assuming potential model 3.  Since for the equilibrium SWNT junction, 
the potential variation is appreciable over a length scale comparable 
to half of the SWNT length and up to $\sim 10 (nm)$, both the magnitude 
and the voltage-dependence of the current will be sensitive to the spatial 
variation of the potential response to the applied voltage over the same 
length scale, which may have different effects on the SWNT band  
structure depending on the metallic electrodes used (Fig. \ref{xueFig17}). 
Therefore accurate modeling of this long-range potential 
variation at the metal-SWNT interface will be critical for evaluating the 
current-transport mechanism of the nanoscale SWNT devices. 
 
\section{Conclusion}

The rapid development of single-wall carbon nanotube-based device technology 
presents opportunities both for exploring novel device concepts based on 
atomic-scale nanoengineering techniques and for examining the physical 
principles of nanoelectronics from the bottom-up atomistic approach. As the 
first example of the device physics problems raised in this context, we examine 
electron transport through metal-SWNT interface when the finite SWNT is 
contacted to the metal surfaces through the dangling bonds at the end, 
which presents an atomic-scale analogue to the planar metal-semiconductor 
interface. Due to the quasi-one-dimensional geometry of the SWNTs, 
a correct understanding of the physical mechanisms involved requires an 
atomistic analysis of the electronic processes in the configuration of the 
metal-SWNT-metal junctions.   
   
We have presented in this paper such a microscopic study of electronic and 
transport properties of metal-SWNT interfaces, as the length of the finite 
SWNT varies from the molecular limit to the bulk limit and the strength 
of the interface coupling varies from the strong coupling to the weak coupling 
limit. Our models are based on a self-consistent tight-binding implementation 
of the recently developed self-consistent matrix Green's function (SCMGF) 
approach for modeling molecular electronic devices, which includes atomistic 
description of the SWNT electronic structure, the three-dimensional 
electrostatics of the metal-SWNT interface and is applicable to arbitrary 
nanostructured devices within the coherent transport regime. We present a 
bottom-up analysis of the nature of the Schottky barrier formation, the length 
and temperature dependence of electron transport through the metal-SWNT 
interfaces, which show quite different behavior compared to the planar 
metal-semiconductor interfaces, due to the confined cylindrical geometry and 
the finite number of conduction channels within the SWNT junctions. 
We find that the current-voltage characteristics of the metal-SWNT-metal 
junctions depend sensitively on the electrostatic potential profiles across the 
SWNT junction, which indicates the importance of the self-consistent modeling 
of the long-range potential variation at the metal-SWNT interface for 
quantitative evaluation of device characteristics. 

Much of current interests on the Schottky barrier effect at metal-SWNT 
interface are stimulated by the controversial role it plays in the operation 
of carbon nanotube field-effect transistors 
(CNTFET),~\cite{AvFET,DaiFET,McEuFET} where different contact schemes 
and metallic electrodes have been used. In general, the operation of 
CNTFET will be determined by the combined gate and source/drain voltage  
effect on the Schottky barrier shape at the metal-SWNT interface, which 
may depend on the details of the metal-SWNT contact geometry, nanotube 
diameter/chirality and temperature/voltage range. Correspondingly, an  
atomic-scale understanding of the gate modulation effect within the 
metal-insulator-SWNT capacitor configuration will also be needed, similar 
to the planar metal-oxide-semiconductor structure.~\cite{MOSBook} 
We believe that detailed knowledges of the electronic processes within 
both the metal-SWNT-metal junction and the metal-insulator-SWNT 
capacitor are needed before a clear and unambiguous picture on the 
physical principles governing the operation of CNTFET can emerge. In 
particular, preliminary theoretical results on the carbon-nanotube 
field-effect transistors  
show that for SWNT molecule end-contacted to the electrodes, the nanotube 
transistor functions through the gate modulation of the Schottky barrier at the 
metal-SWNT interface (in agreement with recent experiments,~\cite{AvFET})  
which becomes more effective as the length of the SWNT molecule 
increases. Further analysis is thus needed that treat both the gate and 
source/drain field self-consistently within the SWNT junctions, 
to achieve a thorough understanding of SWNT-based nanoelectronic devices.   
      
\begin{acknowledgments}
This work was supported by the DARPA Moletronics program, the NASA URETI 
program, and the NSF Nanotechnology Initiative. 
\end{acknowledgments}


\begin{references}
\bibitem[*]{Xue} Author to whom correspondence should be addressed.
\bibitem{Note} The work functions of gold and titanium used here are those 
of polycrystalline materials since the metallic electrodes used in most 
experiments are almost never close to being single-crystalline.
\end{references}
\end{document}